\documentclass[modern]{aastex631}
\usepackage{upgreek}
\begin{document}

\title{Characterization of the ejecta from NASA/DART impact on Dimorphos: observations and Monte Carlo models}

\author[0000-0003-0670-356X]{Fernando Moreno}
\affiliation{Instituto de Astrof\'isica de Andaluc\'ia, CSIC \\
Glorieta de la Astronom\'ia, s/n, 18008 Granada, Spain}

\author{Adriano Campo Bagatin}
\affiliation{
Instituto de F\'isica Aplicada a las Ciencias y las Tecnolog\'ias (IUFACyT),  Universidad de Alicante, \\ San Vicent del Raspeig, 03690 Alicante, Spain}
\affiliation{Departamento de F\'isica, Ingenier\'ia de Sistemas y Teor\'ia de la Se\~nal \\ Universidad de Alicante, San Vicent del Raspeig, 03690 Alicante, Spain}

\author{Gonzalo Tancredi}
\affiliation{Departamento de Astronom\'ia, Facultad de Ciencias, Igu\'a 4225, 11400 Montevideo, Uruguay}

\author{Jian-Yang Li}
\affiliation{Planetary Science Institute, Tucson, AZ, USA}

\author{Alessandro Rossi}
\affiliation{FAC-CNR, Via Madonna del Piano 10, 50142, Sesto Fiorentino, Italy}

\author{Fabio Ferrari}
\affiliation{Department of Aerospace Science and Technology, Politecnico di Milano, Milano, Italy}

\author[0000-0002-1821-5689]{Masatoshi Hirabayashi}
\affiliation{Auburn University, Auburn, AL, USA}
\affiliation{Daniel Guggenheim School of Aerospace Engineering, Georgia Institute of Technology, Atlanta GA 30332, USA}

\author{Eugene Fahnestock}
\affiliation{Jet Propulsion Laboratory, California Institute of Technology, Pasadena, CA, USA}

\author{Alain Maury}
\affiliation{SPACEOBS, San Pedro de Atacama, Chile}

\author{Robert Sandness}
\affiliation{SPACEOBS, San Pedro de Atacama, Chile}

\author{Andrew S. Rivkin}
\affiliation{Johns Hopkins University Applied Physics Laboratory, Laurel, MD, USA}

\author{Andy Cheng}
\affiliation{Johns Hopkins University Applied Physics Laboratory, Laurel, MD, USA}

\author{Tony L. Farnham}
\affiliation{University of Maryland, Department of Astronomy, College Park, MD, USA}

\author{Stefania Soldini}
\affiliation{Department of Mechanical, Materials and Aerospace Engineering, University of Liverpool, Liverpool, UK}
\author{Carmine Giordano}
\affiliation{Department of Aerospace Science and Technology, Politecnico di Milano, Milano, Italy}
\author{Gianmario Merisio}
\affiliation{Department of Aerospace Science and Technology, Politecnico di Milano, Milano, Italy}
\author{Paolo Panicucci}
\affiliation{Department of Aerospace Science and Technology, Politecnico di Milano, Milano, Italy}
\author{Mattia Pugliatti}
\affiliation{Department of Aerospace Science and Technology, Politecnico di Milano, Milano, Italy}

\author{Alberto J. Castro-Tirado}
\affiliation{Instituto de Astrof\'isica de Andaluc\'ia, CSIC \\
  Glorieta de la Astronom\'ia, s/n, 18008 Granada, Spain, and
Unidad Asociada al CSIC, Departamento de Ingenier\'ia de Sistemas y Autom\'atica, Escuela de Ingenier\'ias, Universidad de M\'alaga, M\'alaga, Spain}

\author{Emilio Fern\'andez-Garc\'ia}
\affiliation{Instituto de Astrof\'isica de Andaluc\'ia, CSIC \\
  Glorieta de la Astronom\'ia, s/n, 18008 Granada, Spain}

\author{ignacio P\'erez-Garc\'ia}
\affiliation{Instituto de Astrof\'isica de Andaluc\'ia, CSIC \\
  Glorieta de la Astronom\'ia, s/n, 18008 Granada, Spain}

\author{Stavro Ivanovski}
\affiliation{INAF - Osservatorio Astronomico di Trieste, Via G.B. Tiepolo, 11, Trieste, Italy}

\author{Antti Penttila}
\affiliation{Department of Physics, P.O.Box 64, FI-00014, University of Helsinki, Finland}

\author{Ludmilla Kolokolova}
\affiliation{Department of Astronomy, University of Maryland, College Park, MD, USA}

\author{Javier Licandro}
\affiliation{Instituto de Astrof\'isica de Canarias C/Vía Láctea s/n, 38205 La Laguna, and Departamento de Astrof\'isica, Universidad de La Laguna, 38206 La Laguna, Tenerife, Spain} 

\author{Olga Mu\~noz}
\affiliation{Instituto de Astrof\'isica de Andaluc\'ia, CSIC \\
  Glorieta de la Astronom\'ia, s/n, 18008 Granada, Spain}

\author{Zuri Gray}
\affiliation{Armagh Observatory \& Planetarium, College Hill, Armagh, BT61 9DG, UK, and Mullard Space Science Laboratory, Department of Space and Climate Physics, University College London, Holmbury St. Mary, Dorking, Surrey RH5 6NT, UK}

\author{Jose L. Ortiz}
\affiliation{Instituto de Astrof\'isica de Andaluc\'ia, CSIC \\
  Glorieta de la Astronom\'ia, s/n, 18008 Granada, Spain}

\author{Zhong-Yi Lin}
\affiliation{Institute of Astronomy, National Central University, No. 300, Zhongda Rd., Zhongli Dist., Taoyuan City 32001, Taiwan}




\begin{abstract}

The NASA/DART (Double Asteroid Redirection Test) spacecraft successfully crashed on Dimorphos, the secondary component of the binary (65803) Didymos system. Following the impact, a large dust cloud was released, and a long-lasting dust tail was developed. We have extensively monitored the dust tail from the ground and from the Hubble Space Telescope (HST). We provide a characterization of the ejecta dust properties, i.e., particle size distribution and ejection speeds, ejection geometric parameters, and mass, by combining both observational data sets, and  by using Monte Carlo models of the observed dust tail. The differential size distribution function that best fits the imaging data was a broken power-law, having a power index of --2.5 for particles of r$\le$ 3 mm, and of --3.7 for larger particles. The particles range in sizes from 1 $\upmu$m up to 5 cm. The ejecta is characterized by two components, depending on velocity and ejection direction. The northern component of the double tail, observed since October 8th 2022, might be associated to a secondary ejection event from impacting debris on Didymos, although it is also possible that this feature results from the binary system dynamics alone. The lower limit to the total dust mass ejected is estimated at  $\sim$6$\times$10$^6$ kg, half of this mass being ejected to interplanetary space.

\end{abstract}


\section{Introduction}\label{sec:intro}

The Double Asteroid Redirection Test (DART) is a NASA mission that impacted a spacecraft on the surface of Dimorphos, the satellite of the primary asteroid (65803) Didymos \citep{2018P&SS..157..104C}. On 2022 26th September, 23:14 UT DART impacted in a nearly head-on configuration on Dimorphos surface, giving rise first to a fast ejected material (plume) (speed of $\approx$2 km s$^{-1}$) whose spectrum is consisting of emission lines of ionized alkali metals (NaI, KI, and LiI) \citep{2023Icar..40115595S}. This plume was clearly observed on images obtained from Les Makes Observatory  \citep{2023Natur.616..461G}, right after impact time, and was also seen in the earliest images during the HST monitoring \citep{2023Natur.616..452L}. A wide  ejection cone of dust particles and meter-sized boulders was monitored by the Light Italian CubeSat for Imaging of Asteroid \citep[LICIACube,][]{2021P&SS..19905185D, 2023acm..Farnham} which  performed a fast flyby of the system. 

Apart from the plume, a fraction of the ejected mass was emitted with significantly lower speeds, forming the ejecta pattern and tail seen since the earliest acquired images from ground-based observatories \citep{2023A&A...671L..11O, 2023ApJ...945L..38B}, and from the HST \citep{2023Natur.616..452L}. Our purpose is to characterize the dust properties of this mostly slow-moving ejecta, the ejection velocities, the size distribution, and the ejected mass, using Monte Carlo models to simulate the motion of the particles in the spatial region near the binary system. After describing the ejecta observations in section \ref{sec:obs}, in section \ref{sec:model} we introduce the Monte Carlo models used to calculate the synthetic tail brightness and their time evolution, and discuss the results obtained. Finally, the conclusions are given in Section \ref{sec:conclusions}.      

\section{Observations}\label{sec:obs}

We first describe the observational material acquired from the ground, followed by a brief description of the HST observations \citep{2023Natur.616..452L}. Table \ref{tab:technical} summarizes the technical data of the instrumentation used.
\begin{table*}
    \caption{Technical data of the instrumentation used.}
    \label{tab:technical}
    \centerline{
    \begin{tabular}{llllll}
Telescope & Location & CCD & Camera        & Plate scale         & Filter \\
          & (Latitude; Longitude) &     & field of view & (\arcsec/pixel) & \\
\hline          
HST       &   -- & Marconi & 160\arcsec $\times$ 160\arcsec & 0.04           &  F350LP \\ 	  
SPACEOBS  & 22$^\circ$57\arcmin 09.8\arcsec S; 68$^\circ$10\arcmin 
 48.7\arcsec W & ZWO ASI6200MM Pro & 49\arcmin $\times$ 29\arcmin & 0.54 & Clear \\
BOOTES    & 37$^\circ$05\arcmin 58.2\arcsec N; 06$^\circ$44\arcmin 
 14.9\arcsec W & Andor iXon EMCCD & 16.8\arcmin $\times$ 16.8\arcmin  & 1.97 & Clear \\
\hline
\end{tabular}}
\end{table*}

Aperture photometry of the binary system was performed using the BOOTES-1 telescope. The Burst Observer and Optical Transient Exploring System (BOOTES) is a world-wide robotic telescope network primarily designed to detect and follow gamma ray bursts (GRBs) \citep{2012ASInC...7..313C, 2023arXiv230206565H}. The aperture photometry measurements were performed using BOOTES-1, which is a 0.3-m aperture telescope located in the Estaci\'on de Sondeos Atmosf\'ericos in the Centro de Experimentaci\'on, El Arenosillo, Huelva, Spain.  The aperture size was selected automatically in the range 6-7$\arcsec$, depending on the seeing conditions.  The photometric data were calibrated using standard stars in the Gaia G-band system \citep{2018A&A...617A.138W}. 

\begin{figure*}
\includegraphics[angle=-90,width=\textwidth]{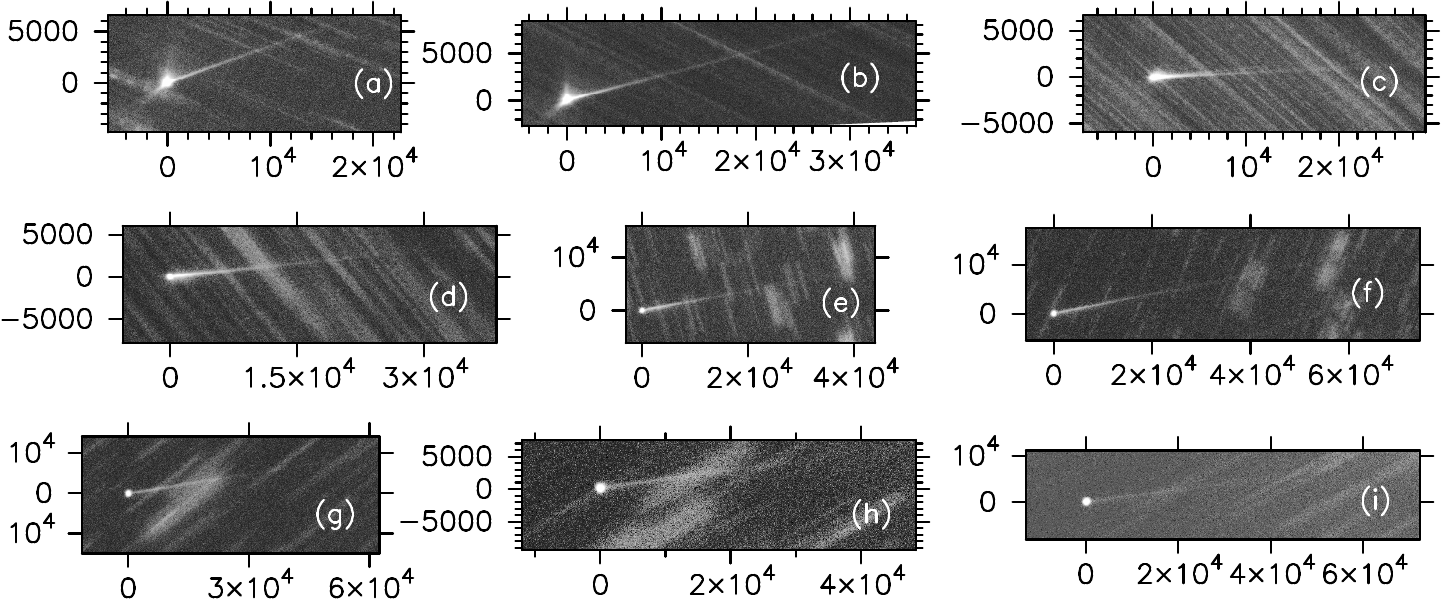}   
\caption{The subset of SPACEOBS images used for modeling. Axes are labeled in km projected distance at the asteroid in all panels. Labels (a) through (i) indicate observation time as given in Table \ref{tab:SPACEOBS}. Celestial North is up, celestial East is left in all panels. 
\label{fig:SPACEOBS-images}}
\end{figure*}

The ground-based images were acquired from a private observatory located in the Atacama desert (Chile) called San Pedro de Atacama Celestial Explorations (SPACEOBS) which is run by Alain Maury. The Atacama desert is an excellent place for astronomical observations, with low humidity, and good transparency and seeing conditions. All the observations were performed with a CCD camera mounted on a 0.43-m aperture telescope. The technical information on the instrumentation used is displayed in Table \ref{tab:technical}. Images were acquired from the impact date (September 26th, 2022) to late December 2022, on 56 epochs in total. The images were acquired using a non-sidereal tracking mode, i.e., by tracking on the binary system, using always an exposure time of 300 s. The reduction of the images was performed by standard techniques, including bias subtraction and flat-fielding. The sky background was estimated in each image by taking a median value of field star-free regions in each frame.  A median image was obtained on each night, by stacking up all the available reduced images. The images were calibrated to magnitudes arcsec$^{-2}$ using the photometric data from BOOTES-1 until October 20th, 2022. At later epochs, we assumed for calibration of the images the $V$ band magnitude values obtained from the JPL Horizons web interface\footnote{  
https://ssd.jpl.nasa.gov/horizons/} for the Didymos system, as the tail contribution is essentially negligible on those dates. This involves the assumption that the "naked" system has not experienced any brightness variation post- to pre-impact conditions, which is confirmed by other observations. Thus, photometric measurements by Pravec et al. (2023, private communication) reveal a difference of just --0.061 mag between pre- and post-impact absolute magnitudes, which has been detected at only 1.9-sigma level (formal errors), so it is only a marginal detection of the binary system's brightening, not statistically significant. In line with this, 
Buratti et al. (2023, personal communication) do not report any significant brightness variation in the system post-impact either, the difference being  of only --0.13 absolute magnitudes relative to pre-impact data. 

\begin{figure}
\includegraphics[angle=-90,width=0.99\columnwidth]{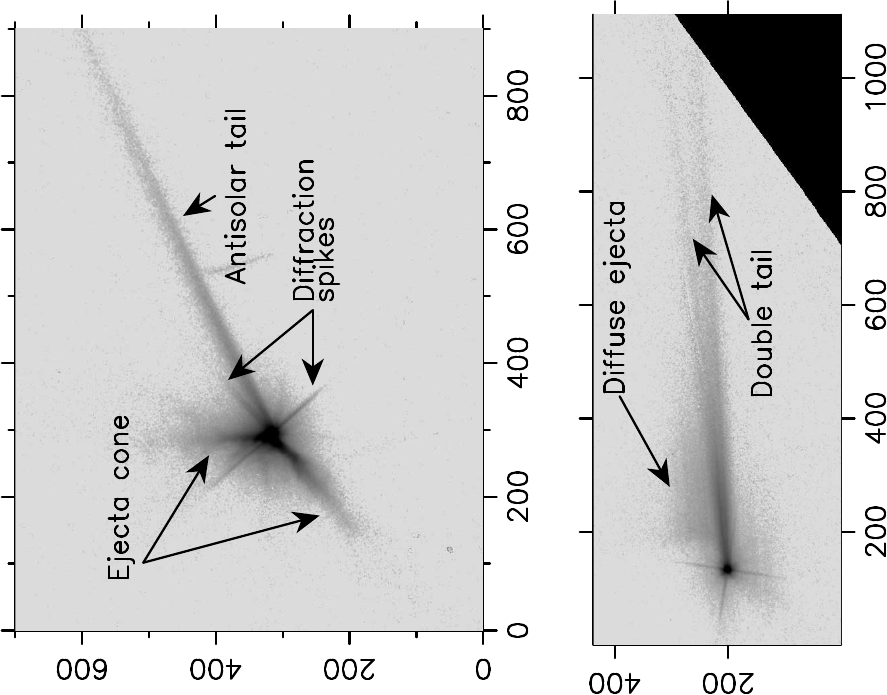}   
\caption{HST images obtained on 2022 28th September (code (l) in Table \ref{tab:HST} (upper panel) and 2022 8th October (code (o) in Table \ref{tab:HST}, indicating the most obvious features encountered in the images. Axes are labeled in pixels, where 1 pixel represents $\approx$2 km on the sky projected. Celestial North is up, celestial East is left in both panels. 
\label{fig:features}}
\end{figure}

From the large observational data set, we selected for modeling those shown in Figure \ref{fig:SPACEOBS-images}, whose observational parameters are given in Table \ref{tab:SPACEOBS}. The earliest images acquired only one day or two after the impact already show a complex morphology, where, in addition to other smaller-scale features, two conspicuous features directed towards north and southeast (the ejecta cone features in Figure \ref{fig:features}, upper panel) became apparent, as well as a well-developed tail in the antisolar direction. In addition, a secondary tail appeared north of the main tail about 6 days after the impact, forming a double tail feature barely seen in the ground-based images (see Figures \ref{fig:doubletail} and \ref{fig:lulin}), but clearly seen in the HST images \citep[see Figure \ref{fig:features}, lower panel, and see also][]{2023Natur.616..452L}. The origin of the northern component of the double tail is still unclear, but it clearly follows the corresponding synchrone at T$_0$+6$\pm$1 day (where T$_0$ is the impact time) \citep{2023Natur.616..452L}.

\begin{table*}
    \caption{Log of SPACEOBS observations.}
    \label{tab:SPACEOBS}
    \centerline{
    \begin{tabular}{cccccccc}

\multicolumn{1}{c}{Time} & \multicolumn{1}{c}{Time since}& \multicolumn{1}{c}{$r_h$\footnote{Heliocentric distance}} & \multicolumn{1}{c}{$\Delta$\footnote{Geocentric distance}}&\multicolumn{1}{c}{PsAng\footnote{Position angle of the extended Sun-to-asteroid radius vector}} & \multicolumn{1}{c}{PlAng\footnote{Angle between observer and asteroid orbital plane}} &\multicolumn{1}{c}{$\alpha$\footnote{Phase angle}} &Code\\
    \multicolumn{1}{c}{(UT)} & \multicolumn{1}{c}{impact (days)} & \multicolumn{1}{c}{(au)} &\multicolumn{1}{c}{(au)} &\multicolumn{1}{c}{(deg)} & \multicolumn{1}{c}{(deg)} & \multicolumn{1}{c}{(deg)} & \\
    \hline
 2022-Sep-30 07:41&  3.32 & 1.0378 &  0.072 &292.45 &  48.67 &  58.24 & (a) \\
 2022-Oct-03 06:43&  6.28 & 1.0315 &  0.071 &288.23 &  48.19 &  62.56 & (b)\\
 2022-Oct-16 07:12& 19.30 & 1.0145 &  0.081 &280.87 &  34.34 &  75.20 & (c) \\
 2022-Nov-02 07:26& 36.31 & 1.0202 &  0.113 &283.72 &  16.63 &  72.70 & (d) \\
 2022-Nov-18 08:24& 52.35 & 1.0533 &  0.147 &284.81 &   6.80 &  60.31 & (e) \\
 2022-Dec-02 07:26& 66.31 & 1.1007 &  0.176 &282.22 &   1.05 &  45.82 & (f) \\
 2022-Dec-17 06:00& 81.25 & 1.1657 &  0.212 &273.76 &  -3.30 &  28.33 & (g)\\ 
 2022-Dec-22 08:10& 86.34 & 1.1904 &  0.227 &268.42 &  -4.38 &  22.32 & (h) \\
 2022-Dec-24 07:13& 88.30 & 1.2001 &  0.233 &265.75 &  -4.73 &  20.06 & (i) \\
\hline
    \end{tabular}}
\end{table*}

The HST images, already described in \cite{2023Natur.616..452L}, were acquired using the 2.4-m diameter HST with the Wide Field Camera 3 (WFC3). Additional technical details of the instrumentation used are provided in Table \ref{tab:technical}. We have selected for modeling a subset of the HST calibrated images as shown in Table \ref{tab:HST}. Images coded as (l) and (o) are depicted in Figure \ref{fig:features}, showing the most conspicuous features observed in the images, providing a nomenclature reference. 

\begin{table*}
    \caption{Log of the HST observations.}
    \label{tab:HST}
    \centerline{
    \begin{tabular}{cccccccc}

\multicolumn{1}{c}{Time} & \multicolumn{1}{c}{Time since}& \multicolumn{1}{c}{$r_h$} & \multicolumn{1}{c}{$\Delta$}&\multicolumn{1}{c}{PsAng} & \multicolumn{1}{c}{PlAng} &\multicolumn{1}{c}{$\alpha$} &\multicolumn{1}{c}{Code} \\
    \multicolumn{1}{c}{(UT)} & \multicolumn{1}{c}{impact (days)} & \multicolumn{1}{c}{(au)} &\multicolumn{1}{c}{(au)} &\multicolumn{1}{c}{(deg)} & \multicolumn{1}{c}{(deg)} & \multicolumn{1}{c}{(deg)} &  \\
    \hline
 2022-Sep-27 01:04 & 0.04 &   1.046   &  0.076  &  297.84   & 47.59   &  53.34 & (j) \\
 2022-Sep-27 07:25 &  0.31 &   1.045   &   0.075  &  297.39   & 47.73 &  53.74 & (k) \\
 2022-Sep-28 02:28 &  1.10 &    1.043   &   0.074  &  296.05   & 48.11  &  54.92 & (l)\\
 2022-Oct-01 16:12 &  4.67  &    1.035   &   0.072  &  290.42  &  48.63  &  60.25 & (m)\\
 2022-Oct-05 18:38 &  8.78  &   1.027   &  0.071   & 285.41   & 46.68   & 65.94 & (n)\\
 2022-Oct-08 19:40 & 11.82 &  1.022   &  0.073   & 282.95  &  43.76  &  69.56 & (o)\\
 2022-Oct-11 20:42 & 14.86  & 1.018   &  0.075  & 281.53  &  40.10  &  72.44 & (p)\\
 2022-Oct-15 10:26 & 18.43  & 1.015   &  0.078   & 280.90   & 35.46  &  74.80 & (q)\\
\hline
    \end{tabular}}
\end{table*}

In order to refer all the aperture photometry data and images to a common photometric system, the solar spectrum in combination with the reflectance spectrum of the binary system should be taken into account. The output of our Monte Carlo codes is given in solar disk intensity units ($i/i_\odot$), that we converted to $r^\prime$ Sloan mag arcsec$^{-2}$, $m$, to compare with the observations, according to the equation:
\begin{equation}
 m = 2.5\log_{10}\Omega + m_\odot -2.5\log_{10} (i/i_\odot)
     \label{eqsolardisk}
\end{equation}
where $\Omega$ is the solid angle subtended by the Sun at 1 au expressed in arcsec$^2$ ($\Omega$=2.893$\times 10^6$ arcsec$^2$), and $m_\odot$ is the magnitude of the Sun in the $r^\prime$ Sloan filter, $m_\odot$=--26.95 
\citep{2001AJ....122.2749I}. 

If the reflected spectrum were purely solar, the aperture photometry data, given in the G-band system, could be converted to $r^\prime$ by $r^\prime$=$G$+0.066 mag \citep{2017EPSC...11..743O}. On the other hand, the conversion of magnitudes in the HST \texttt{F350LP} filter to Johnson's $V$ has been given by \cite{2019GeoRL..46.1956N} as $V$= $\texttt{F350LP}$--0.12 mag. Then, using the relation  $r^\prime = V-0.49(B-V)+0.11$ mag \citep{1996AJ....111.1748F}, valid for stars with $(B-V) \le +1.5$ mag, and the solar color index $(B-V)$=0.629 mag \citep{2018ApJS..236...47W} we get $r^\prime$=$\texttt{F350LP}$--0.32 mag. However, the reflectance spectrum of the unaltered 
Didymos-Dimorphos system exhibits temporal variations in slope that can be attributed to a number of reasons, including
compositional changes on Didymos surface \citep{2022PSJ.....3..183I}, preventing us to perform any precise photometric correction. In addition, the spectrum of the freshly ejected material after DART collision might be spectroscopically different as well. Then, we decided to maintain all the measurements in their original units. In any case, based on the given color index conversion assuming a solar-like spectrum, we do not expect variations higher than $\approx$0.3 mag among the different bands (Gaia G and $\texttt{F350LP}$) and the $r^\prime$ Sloan magnitudes.

\section{Dust tail modeling}\label{sec:model}

Our purpose is to perform an interpretation of the available observed images by Monte Carlo techniques, i.e., by direct calculation of the orbits of the individual particles ejected at the time of impact, and the computation of their positions in space at the time of the observation. 

To calculate the orbits of the ejected dust particles, we used two different approaches. The first one, which we call simple Monte Carlo modeling, assumes that the particles are initially placed out of the Hill sphere of the system, where the gravity of the binary components can be neglected, and then the dust grains are influenced by solar gravity and radiation pressure forces only. In consequence, the particle undergo purely Keplerian orbits around the Sun, and their orbits can be easily integrated. This approach is adequate to describe the escaping ejecta, and is suited to analyze the large-scale, low-resolution, ground-based SPACEOBS observations. The second approach, which we call detailed Monte Carlo model, is used to describe the dynamics in the innermost region close to the binary system, and is based on the integration of the equations of motion of the particles taking fully into account the gravitational fields of the two bodies. Owing to the superb spatial resolution of the HST images, this approach is better suited to analyze the details of the features that appear in those images, but has the obvious drawback of the large CPU time needed to run the model in comparison with the simple Monte Carlo model.  

After setting the particle scattering properties, particularly the geometric albedo, the results of the models (i.e., the evolution of the tail brightness with time) depend on three basic parameters: the ejection velocities, the size distribution, and the total dust mass ejected after the DART impact. The combination of those parameters affect the tail brightness in an intricate  manner. Thus, small dust particles are highly affected by radiation pressure, and quickly populate the far tail regions, while larger particles need a much longer time to leave the near-nucleus\footnote{In the context of this paper, we always refer to "nucleus" as to the two asteroids of the binary system, that are seen as an effective single body in the comet-like  appearance in the Earth-based telescope images of the ejecta} region, depending on ejection speed. In addition, in the detailed model calculations, those particles might be trapped for a long time orbiting close to the binary system and leaking out from it very slowly, mainly if the ejection speeds are close to the escape velocity of Dimorphos. High ejection speeds spread out the particles quickly, so that the tail brightness will tend to decrease.  The size distribution, which is commonly set to a power-law function, defines the range of sizes that dominate the total mass. Thus, for power exponents lower than --4 most of the mass would be concentrated in the smallest particles, while for exponents higher than --3 most of the mass would reside on the larger ones.

The fitting procedure is based on selecting upper and lower limits to the parameter inputs, and experiment with them until a reasonable agreement with all the observations is achieved. Due to the many parameters involved, we cannot assure that the best-fitting parameters constitute the only solution to the problem. The main weakness of the modeling resides in the difficulty of constraining the total mass ejected,  on one hand because of the presence in the particle population of meter-sized and larger boulders, that contribute mostly to the mass, but not to the brightness, when compared with the much more abundant small particle population, and, on the other hand, to the very high-speed ejecta \citep[see e.g.]{2023acm..Fitzsimmons} immediately after impact, which leaves the field of view of the cameras in a very short time interval. We will return back to these problems in the next Section.

\subsection{Simple Monte Carlo modeling}\label{subsec:classMC}

The interpretation of the ground-based dust tail brightness in terms of the simple dynamical-radiative models is made using our Monte Carlo model as described in e.g. \cite{2022MNRAS.515.2178M} and references therein. In such approach, as stated above, the particles are assumed to be affected by the solar gravity and solar radiation pressure only, ignoring the gravity perturbations of the two components of the binary system. Then, this model is valid out of the Hill sphere of the system, i.e., it is useful to characterize the material that has gravitationally escaped from the binary system, but cannot be used to describe the complex dynamics in the vicinity of the asteroid pair. In fact, we will see with the detailed Monte Carlo model that a significant fraction of the ejected mass is lost in collisions with either Didymos or Dimorphos, thus reducing the dust mass ejected to interplanetary space. 

In the simulations, a large amount ($\gtrsim 10^7$) of particles are released with a certain velocity distribution and particle size distribution. The total ejected mass must be also specified. For this application of the code, all the particles are assumed to be ejected instantly, except for a secondary ejection event occurring a few days after the impact, to explain the development of an additional tail component forming a small angle with the main tail, and north of it, to be described at the end of this section.
The particles are considered spherical, independent scatterers, and they do not experience neither collisions among them nor disruption or fragmentation phenomena. Their dynamics is governed by the so-called $\beta$ parameter (not to be confused with the momentum transfer efficiency due to the DART impact, usually denoted also by $\beta$), defined as the
ratio of solar radiation pressure force to solar gravity force, as    
$\beta = F_{rad}/F_{grav} = C_{pr}Q_{pr}/(2\rho_p r)$. In that equation, 
$r$ is the particle radius and $\rho_p$ its density (assumed at 3500 kg m$^{-3}$), $C_{pr}$ = 1.19 $\times$ 10$^{-3}$ kg m$^{-2}$ is the radiation pressure constant, $Q_{pr}$ is the
scattering efficiency for radiation pressure, which becomes 
$Q_{pr}\approx$1 for moderately absorbing particles with $r \gtrsim$1 $\upmu$m \citep[see, e.g.][their Figure 5]{2012ApJ...752..136M}. The assumed density of $\rho_p$=3500 kg m$^{-3}$ corresponds to the density of ordinary chondrite meteorites associated to the S-type spectrum exhibited by the Didymos-Dimorphos system \citep{2013Icar..222..273D}. All particles are assumed as having the same density. The Keplerian trajectories of the particles can be determined from their $\beta$ parameter and the ejection velocity vector. At the end of the integration time, their positions on the sky plane at any time after ejection are recorded.  
The brightness contribution of each particle in a given pixel of the synthetic image, $m$, expressed in mag arcsec$^{-2}$, is given by:
\begin{equation}
 p_R\pi r^2 = \frac{2.24 \times 10^{22} \pi r_h^2\Delta^2 10^{0.4
     (m_{_\odot}-m)}}
     {G(\alpha)}
     \label{eqbright}
\end{equation}
\noindent
where $r_h$ is the asteroid heliocentric distance in au, $\Delta$ is the geocentric distance of the asteroid, and $m_{\odot}$ is the apparent solar magnitude in the appropriate passband. The particle's geometric albedo at zero phase angle is given by $p_R$, and $G(\alpha) = 10^{-0.4\alpha\phi}$ is the phase correction, where $\alpha$ is the phase angle, and $\phi$ is the linear phase coefficient. Recent work by \cite{2023PSJ.....4...24L}, however, shows geometric albedo dependence with phase angle, revealing values between 0.07 and 0.15  for $p_R$ for a range of particle sizes, compositions, and different porosities from several sources, including laboratory data by \cite{2020ApJS..247...19M} and emitted particles from asteroid Bennu \citep{2020JGRE..12506381H}, for phase angles smaller than about 60$^\circ$, so that we adopted $p_R$=0.1 and $G(\alpha)$=1. In any case, in the geometric optics approximation, which holds for the derived size distribution functions, the ejected mass is inversely proportional to the geometric albedo, so that for higher albedoes the dust mass ejected would be lower accordingly. 
In the vicinity of the image optocenters, the contribution of the nucleus reflected light (i.e., the scattered light of the spherical body having an equivalent radius to the Didymos+Dimorphos system), is important, as it may be comparable, or higher than, the dust cloud brightness. In fact, for images taken a few weeks after impact, the contribution of the nucleus brightness to the total brightness is dominant. The equivalent radius of the system can be approximately computed as an average of the Didymos radius and the effective radius of Didymos+Dimorphos system which turns out to be $R_n$=395 m, i.e., only a bit higher than Didymos radius as Didymos has a much larger surface than Dimorphos.  Then, to compute the contribution of the nucleus we assume such spherical body with the same value of geometric albedo given above for the particles. Following the magnitude--phase relationship by \cite{1997SoSyR..31..219S}, for \mbox{$p_R$ = 0.1}, we get $\phi=0.013-0.0104\ln p_R$=0.037 mag deg$^{-1}$. This value is very close to that obtained by Buratti et al. (2023, personal communication) of $\phi$=0.035$\pm$0.001 mag deg$^{-1}$.

The ejection of material is modeled mainly by two ejecta components, traveling at different speeds. This is justified below in order to reproduce both the antisolar tail (slow-speed component), and the conical features (high-speed component). This component, which contributed with 1/3 to the total ejected mass,  is assumed to be characterized by a hollow conical shape whose axis is oriented to J2000 equatorial coordinates RA=130$^\circ$, DEC=17$^\circ$, which is within the range of the current determinations. The impactor direction was RA=128$^\circ$, DEC=18$^\circ$ \citep[e.g.][]{2023acm..Hirabayashi}. Detailed recent calculations of the ejecta geometry by \cite{2023acm..Hirabayashi} predict an emission cone elongated along the north-south direction of Dimorphos with cone axis oriented to RA=140$\pm$4$^\circ$, DEC=17$\pm$7$^\circ$ (uncertainties are 1-$\sigma$ values). However, the precise axis direction does not have a significant impact on the results as long as it does not deviate by more that 10$^\circ$ from the assumed direction. The cone aperture is set to 140$^\circ$, and the cone wall thickness is set to 10$^\circ$. The second ejecta component is described by a hemispherical ejection with the same axis as the conical emission, and contributing with 2/3 to the ejected mass.  

The remaining model parameters are the size distribution and the initial speeds. The size distribution function is initially set to a single differential power-law distribution function with power exponent $\kappa$, i.e., $dn \propto r^{\kappa} dr$, where $dn$ is the number of particles between $r$ and $r+dr$. We assumed an initial value for $\kappa$ as $\kappa$=--2.5, close to the value obtained by \cite{2023Natur.616..452L} on the earliest HST images. The size distribution was assumed to be the same for all the ejecta components. 

Concerning ejection velocities, conventional scaling laws for cratering ejecta generally refer to velocity distributions as a function of launch position 
\citep[e.g.][]{1999M&PS...34..605C, 1983JGR....88.2485H, 2011Icar..211..856H}, and do not include the effects of different sizes of the particles populating the distribution. Only a few experimental or observational studies provide information on velocity distribution as a function of grain size, but, in all cases, because of technical limitations, they refer to sizes in the mm range and larger, up to boulder-sized debris \citep[e.g.][]{2022Icar..38715212O}. After repeated experimentation with the model, it soon became apparent the need of a double component for the ejecta speeds, one associated to faster particles giving rise to the two features associated to the conical ejection (high-speed component), and another, slow-speed component, with ejection velocities close to Dimorphos escape velocity to properly model the length and thickness of the antisolar tail (the hemispherical ejecta component). The faster ejecta was modeled following a power-law function of the particle size, as it has been set to model the ejection speeds for natural impacts on asteroids  (596) Scheila \citep{2011ApJ...740L..11I} and 354P/LINEAR  \citep{2013A&A...549A..13K,2017AJ....153..228K}. On the other hand, the velocities of the slow-speed component are modeled as  $v=0.05(1+\chi)$ m s$^{-1}$, being $\chi$ a random number in the $(0,1)$ interval. The randomization in the speed distribution is imposed in an attempt to mimic somehow its stochastic nature. 
The high-speed ejecta was modeled by $v=0.375 \chi r^{-0.5}$ m s$^{-1}$ (with $r$ expressed in m). Ejecta speed estimates of $\sim$2 m s$^{-1}$ for mm-sized particles have been reported by \cite{2023arXiv230605908R} from ALMA observations of the DART impact. This is in line with our average higher speed ejecta estimates of $\sim$6 m s$^{-1}$ for $r$=1 mm particles. Having a double ejecta component is motivated by the appearance of the conical feature in combination with the antisolar tail: this tail cannot be modeled assuming the conical high-speed component as it would generate a tail far broader and diffuse than observed. The slow-speed component, which encompasses most of the ejected mass (2/3 of the total dust mass) could be associated to the large amount of material which is ejected at a slow velocity during the latter stages of crater formation, as determined from conventional scaling laws \citep[e.g.][]{1983JGR....88.2485H}. In addition, there is another mechanism that might be contributing to this slow-ejecta component, which is the lofting of particles owing to the propagation of seismic waves after the impact \citep{2022MNRAS.tmp.3047T}. 

In conjunction with the two components of the ejecta just described, a third dust ejecta emission event took place on October 2.5, 2022, leading to the secondary, northern branch, of the tail. This event is associated to the presence of the northern secondary tail, that follows the corresponding synchrone at the given epoch. This agrees with the timing obtained by  \cite{2023Natur.616..452L} from HST images (T$_0$+6$\pm$1 days). The small tail thickness suggests low ejection velocities, and its faintness compared to the main tail, a much smaller ejected mass than the main tail slow-component of the ejecta. For simplicity, we adopt the same parameters of the slow ejecta component mentioned above (i.e., $v=0.05(1+\chi)$ m s$^{-1}$), and isotropic ejection.  We will link this dust emission event to impacts of debris particles on Didymos in the framework of the detailed Monte Carlo approach (see section \ref{subsec:detailMC}). At this point, it is interesting to note that a slight increase in brightness has been observed around 6 to 9 days after impact, in both ground-based and HST photometric lightcurves (Kareta et al. 2023, submitted). This so-called "8th-day bump" could be associated to reimpacting material on Didymos, as we will also show later in Section \ref{subsec:detailMC}.

The dust masses ejected for each component that better fits the tail profiles were 2.2$\times$10$^7$ (slow-speed), 7.4$\times$10$^6$ (high-speed), and 3.7$\times$10$^6$ kg (late event), respectively, giving a total mass ejected of 3.3$\times$10$^7$ kg.  The mass of the late ejecta component is just a rough estimate: the signal-to-noise ratio of that secondary tail is too low to allow for a better constraint. This estimate is to be improved with the analysis of the much higher resolution HST images (section \ref{subsec:detailMC}). 

The maximum particle size in the distribution was constrained by the analysis of the latest images. Thus, the initial radius $r_{max}$=1 cm assumed had to be increased to larger values. The reason was that for $r_{max}$=1 cm, the central condensation containing the nuclei would be detached from the tail at the latest observation dates because the radiation pressure would be moving away those $r$=1-cm particles after some months since impact. Then, a larger size limit of $r_{max}$=5 cm was set instead, providing a better fit to the near nucleus region. Regarding the minimum particle size, setting $r_{min}$=1 $\upmu$m was found adequate to fit the outermost part 
of the tail in the earliest images. Also, as we will describe later in section 3, this lower limit is very well constrained by the 
earliest HST images, where the observed length of the tail is very consistent with that minimum size.

\begin{table*}
    \centering
    \caption{Parameters of the best-fit models.}
    \label{tab:models}
    \begin{tabular}{|cccc|c|}
    \hline
    Ejecta & Speed & Ejected  & Ejection & Total unbound\footnote{Delivered to interplanetary space. Note that in the case of the detailed dynamical Monte Carlo model, this mass is not the sum of the total masses ejected due to intervening dynamical stirring and collision of a sizable fraction of the ejecta with Didymos and Dimorphos.}\\
    component & (m s$^{-1}$) & mass (kg) & mode & ejected mass (kg)  \\ \hline
    \multicolumn{4}{|c|}{Simple Monte Carlo model} & \\ \hline
    Slow & 0.05(1+$\xi$) & 2.8$\times$10$^6$ & Hemispherical & \\
     Fast & 0.375$\chi r^{-0.5}$ & 9.2$\times$10$^5$ & Conical &  4.2$\times$10$^6$ \\ 
     Late & 0.05(1+$\xi$) & 4.6$\times$10$^5$ & Isotropic & \\
     \hline
     \multicolumn{4}{|c|}{Detailed dynamical Monte Carlo model} & \\ \hline
   Slow & 0.09 & 4.3$\times$10$^6$ & Hemispherical & \\
     Fast & 0.225$\chi r^{-0.5}$ & 2.1$\times$10$^6$ & Conical & 4.9$\times$10$^6$\\ 
     Late & 0.09 & 3.0$\times$10$^6$ & Isotropic & \\
     \hline
     \end{tabular}
     \end{table*}

\begin{figure}
\includegraphics[angle=-90,width=0.99\columnwidth]{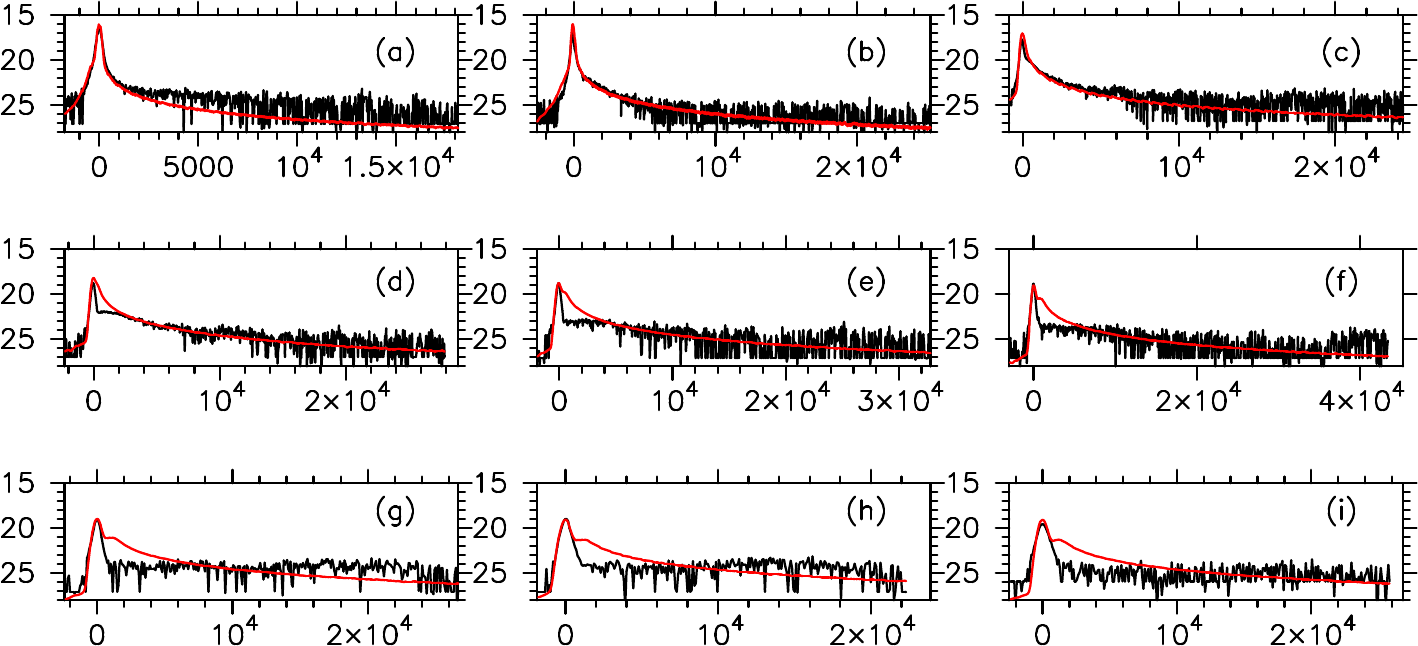}  
\caption{Scans along the tails of the SPACEOBS images. The panels are labeled (a) to (i), corresponding to the dates shown in Table \ref{tab:SPACEOBS} (Code column). The black lines correspond to the observations, the red line to the model. Horizontal axes are 
 labeled in km projected on the sky plane, and vertical axes are expressed in mag arcsec$^{-2}$. These scans were obtained using a single power-law size distribution with $\kappa=-2.5$. The total dust mass released is 3.2$\times$10$^7$ kg. 
\label{fig:singlepower}}
\end{figure}

\begin{figure}
\includegraphics[angle=-90,width=0.99\columnwidth]{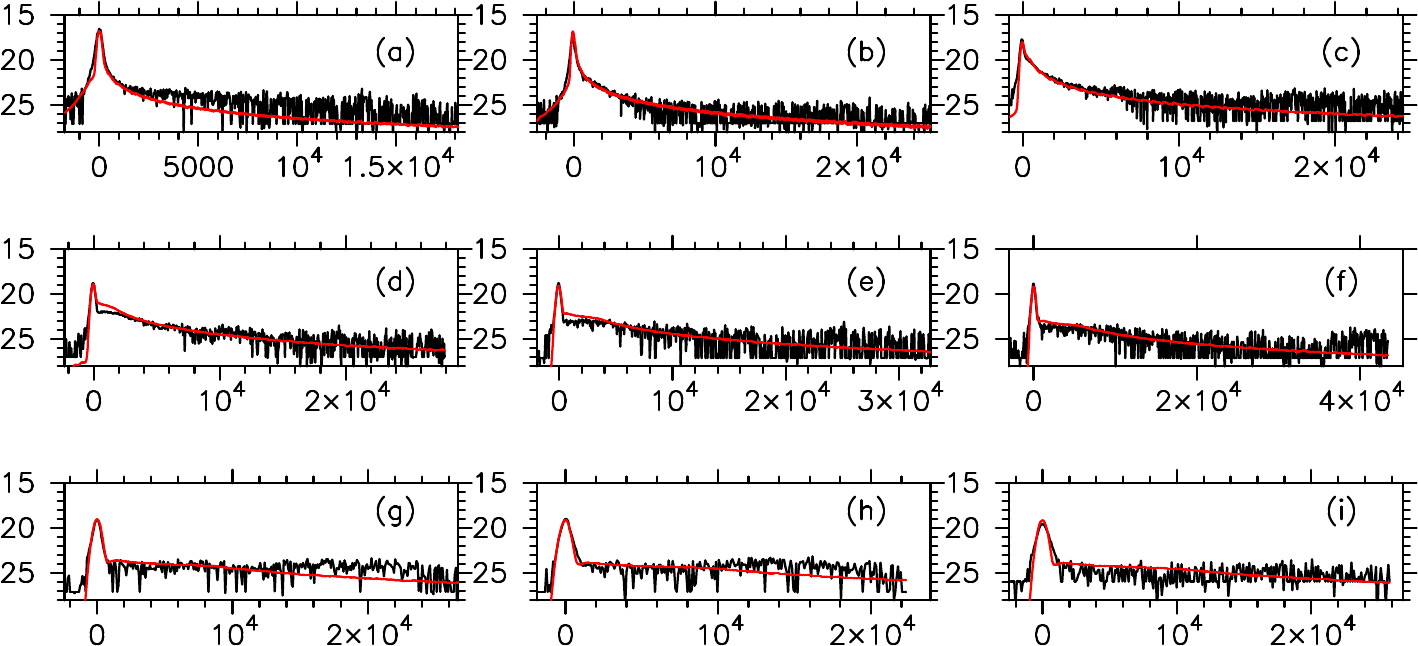}   
\caption{Scans along the tails of the SPACEOBS images. The panels are labeled (a) to (i), corresponding to the dates shown in Table \ref{tab:SPACEOBS} (Code column). The black lines correspond to the observations, the red line to the model. Horizontal axes are 
 labeled in km projected on the sky plane, 
 and vertical axes are expressed in mag arcsec$^{-2}$. These scans were obtained using a broken power-law differential size distribution function with $\kappa=-2.5$ between 1 $\upmu$m and 3 mm, and $\kappa=-3.7$ between 3 mm and 5 cm. The total dust mass sent to interplanetary space is 4.2$\times$10$^6$ kg. 
\label{fig:brokenpower}}
\end{figure}

     With all of the above model inputs, the resulting photometric scans along the tails of the images in comparison with the observations, at the dates shown in Table \ref{tab:SPACEOBS}, are displayed in Figure \ref{fig:singlepower}. Although the fits to the early images are reasonably good, the model does not perform well for images acquired later than $\approx$10 days after impact. Varying the power index $\kappa$ does not produce any improvement either. As the observed brightness in the near-nucleus region is clearly overestimated with this model, we imposed a broken power-law with a ``knee" in the mm size range, to search for an improvement in the fits. We found that a broken power-law with 
$\kappa$=--2.5 for particles smaller than 3 mm in radius, and with a higher slope of $\kappa$=--3.7 for particles having radii larger than 3 mm produce much better fits at all epochs, as can be seen in Figure  \ref{fig:brokenpower}. The assumption of a different size distribution, with a higher slope on the largest particles, implies a recalculation of the ejected masses, that now become a factor of $\approx$8 smaller, i.e., 2.8$\times$10$^6$ kg and 9.2$\times$10$^5$ kg for the slow and fast components, respectively, and of 4.6$\times$10$^5$ kg for the secondary, late, ejecta, giving a total mass of 4.2$\times$10$^6$ kg. It is important to realise that this dust mass constitutes a stringent lower limit to the total ejected mass from Dimorphos. On one hand, the high-speed ($\approx$2 km s$^{-1}$) material released right after the impact \citep[see][]{2023Icar..40115595S} is out of the field of view on our images. On the other hand, the presence in the particle distribution of very large particles, such as boulders, might contribute significantly to the total ejected mass but very little to the brightness, becoming almost undetectable in the images. In that respect, it is convenient to mention the findings by \cite{2023acm..Farnham}, who have detected a boulder population after DART impact by analyzing LICIACube LUKE images. Those authors found a population of some 100 meter-sized boulders, so that, assuming a density of 3500 kg m$^{-3}$, they would give a total mass of $\approx$1.5$\times$10$^6$ kg. Those boulders are moving at speeds of 20-50 m s$^{-1}$, so that they carry a momentum that might be comparable to that of DART spacecraft \citep{2023acm..Farnham}. Let us assume that the actual boulder population was a factor of 100 higher, i.e., a total mass of 10$^8$ kg, and that this population is distributed following a power law of index --3.7 (as in our model), with ejection speeds of 20 m s$^{-1}$. This would result in a unrealistic momentum balance, but we would like to remark that even in this case the boulder population would add a negligible increase in the integrated flux of only 0.06\% relative to the corresponding model results at the dates shown in Table 2. Even if we reduce the speed of those boulders to the much smaller speeds used in the modeling (see Table 4), that will tend to concentrate the boulders much closer to the optocenter at all epochs, the contribution to the integrated flux would be of only 7\% relative to the model results.

\begin{figure}[h]
\includegraphics[angle=-90,width=0.99\columnwidth]{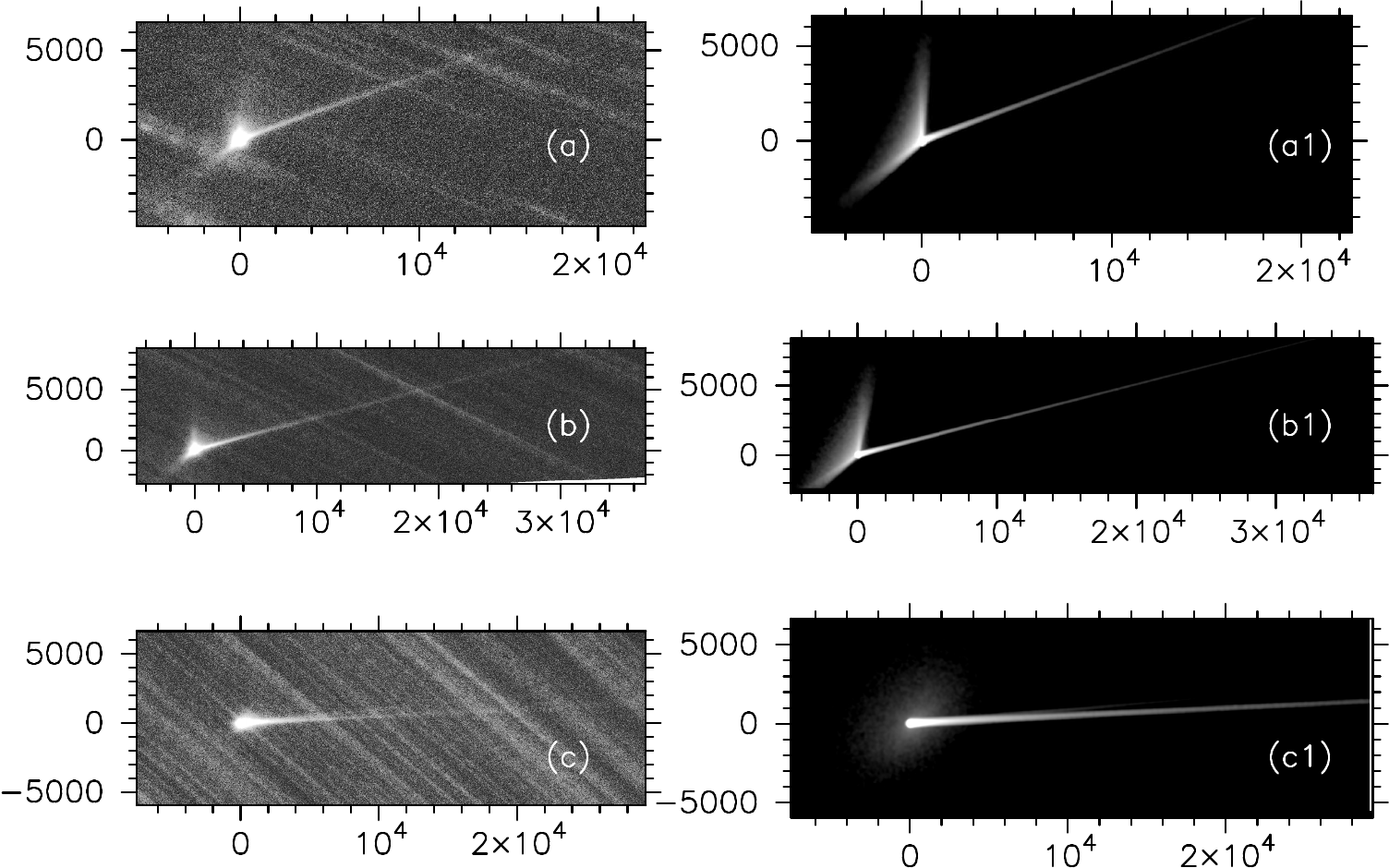}   
\caption{Panels (a), (b), and (c) display the SPACEOBS images at the corresponding dates in Table \ref{tab:SPACEOBS}, and  panels (a1), (b1), and (c1) the corresponding synthetic images generated with the simple Monte Carlo model. All images are stretched between 28 and 22 mag arcsec$^{-2}$. Axes are 
 labeled in km projected on the sky plane. North is up, East to the left in all images.
\label{fig:spaceobs1}}
\end{figure}

\begin{figure}[h]
\includegraphics[angle=-90,width=0.99\columnwidth]{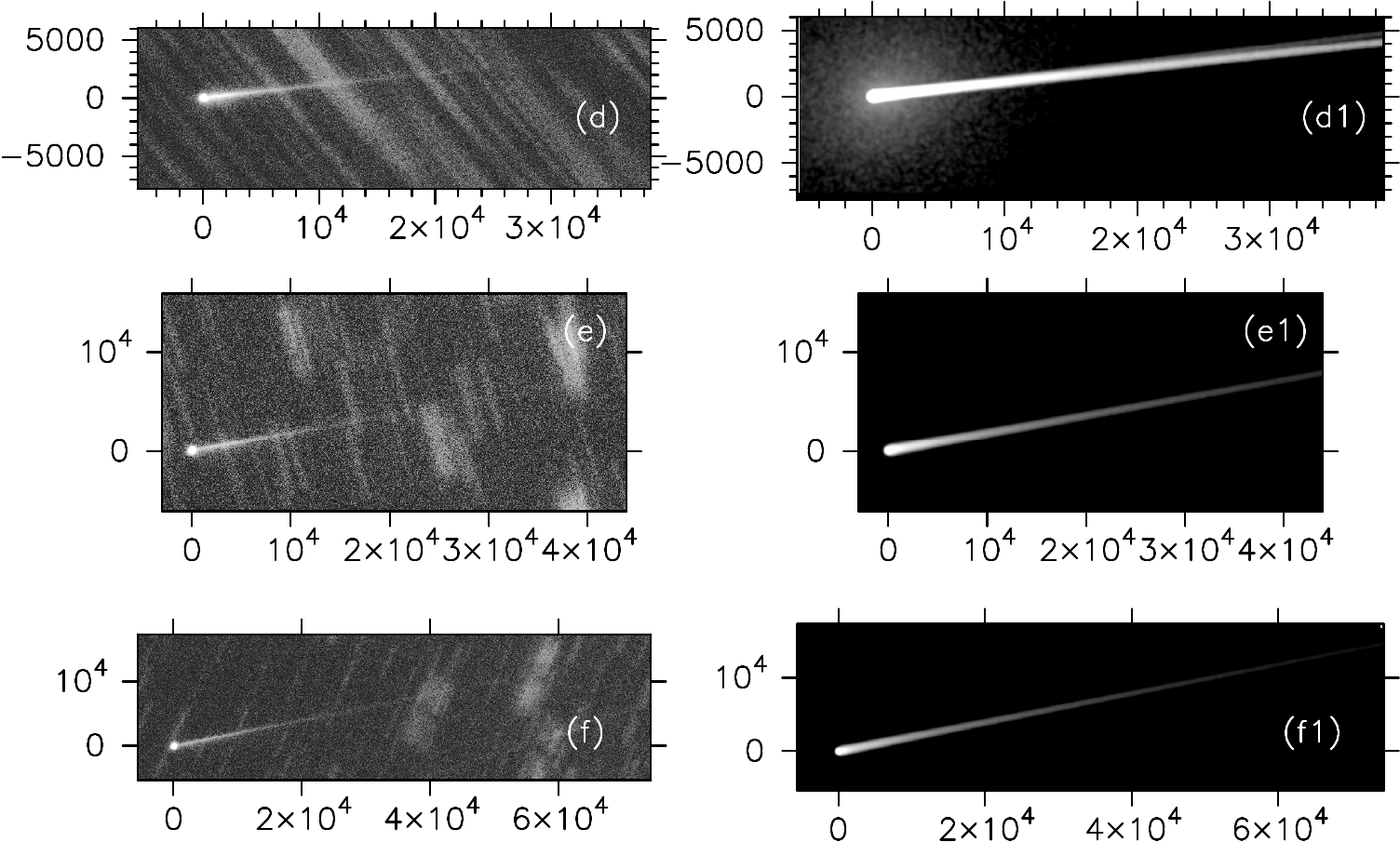}   
\caption{Panels (d), (e), and (f) display the SPACEOBS images at the corresponding dates in Table \ref{tab:SPACEOBS}, and  panels (d1), (e1), and (f1) the corresponding synthetic images generated with the simple Monte Carlo model. All images are stretched between 28 and 22 mag arcsec$^{-2}$, except synthetic image (d1) that has been heavily stretched between 30 and 25 mag arcsec$^{-2}$, barely showing the secondary tail north of the main tail. Axes are 
 labeled in km projected on the sky plane. North is up, East to the left in all images.
\label{fig:spaceobs2}}
\end{figure}

\begin{figure}[h]
\includegraphics[angle=-90,width=0.99\columnwidth]{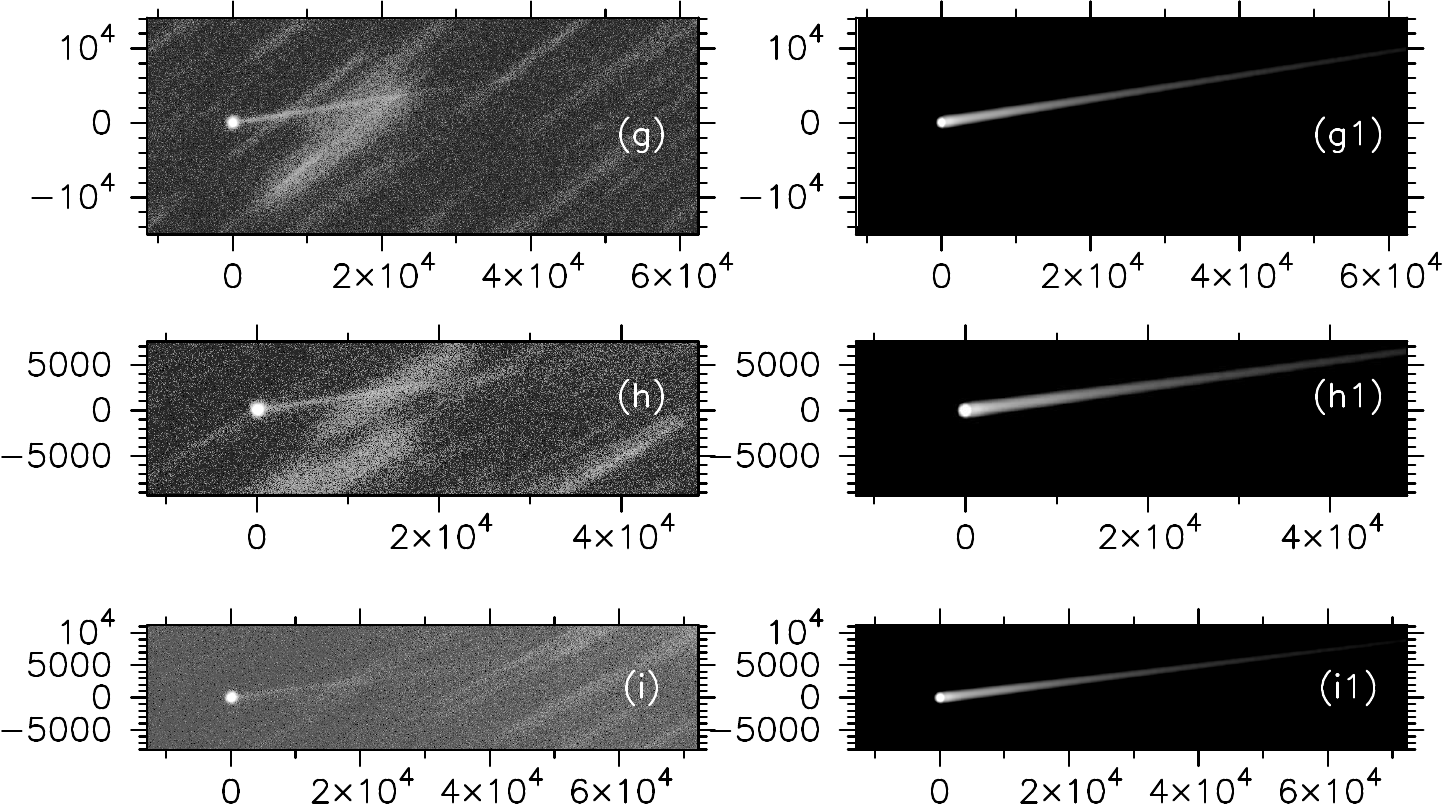}   
\caption{Panels (g), (h), and (i) display the SPACEOBS images at the corresponding dates in Table \ref{tab:SPACEOBS}, and  panels (g1), (h1), and (i1) the corresponding synthetic images generated with the simple Monte Carlo model. All images are stretched between 28 and 22 mag arcsec$^{-2}$. Axes are 
 labeled in km projected on the sky plane. North is up, East to the left in all images.
\label{fig:spaceobs3}}
\end{figure}

The synthetic images generated are convolved with a Gaussian function of full-width at half-maximum consistent with the average seeing point-spread function. The modeled images are then compared to the observed images in Figures \ref{fig:spaceobs1}, \ref{fig:spaceobs2}, and \ref{fig:spaceobs3}, using the same grayscale. As shown, the modelled images capture well many of the features displayed in the observed images.  

\begin{figure}[h]
\includegraphics[angle=-90,width=0.99\columnwidth]{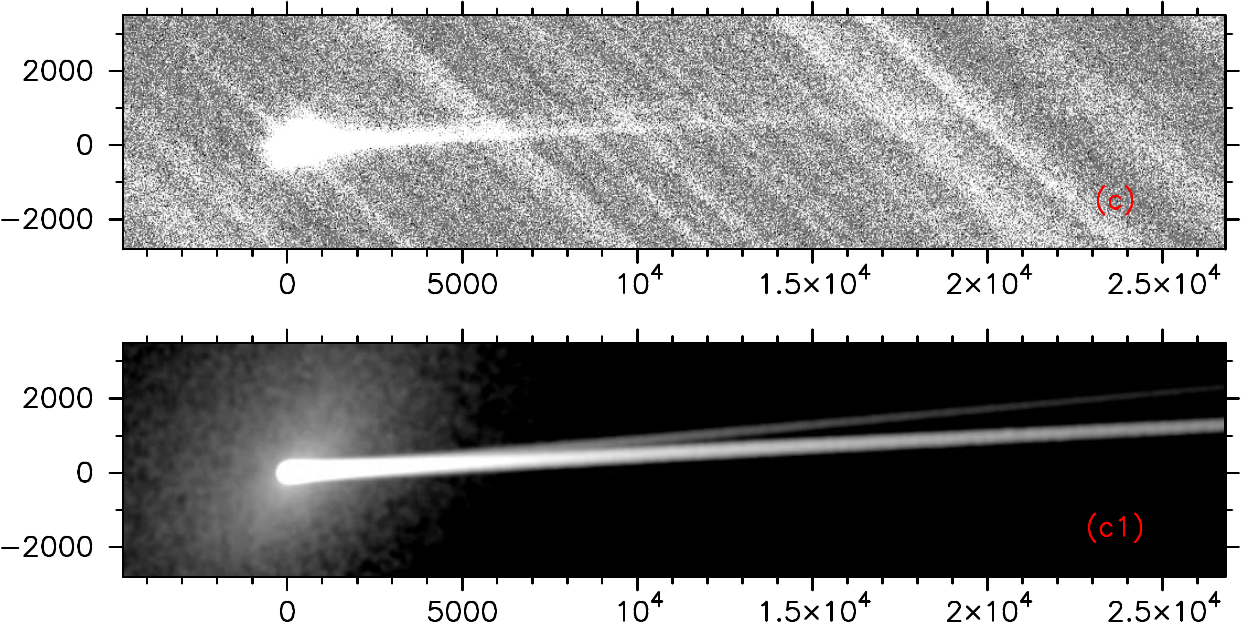}   
\caption{Upper panel: SPACEOBS observation on October 16th, 2022 (image labeled as (c) in Table \ref{tab:SPACEOBS}) of the ejecta showing barely the double tail structure. Lower panel (c1) is the result of the simple Monte Carlo modeling. Axes are labeled in km projected on the sky plane. North is up, East to the left. 
\label{fig:doubletail}}
\end{figure}

\begin{figure}[h]
\includegraphics[angle=-90,width=0.99\columnwidth]{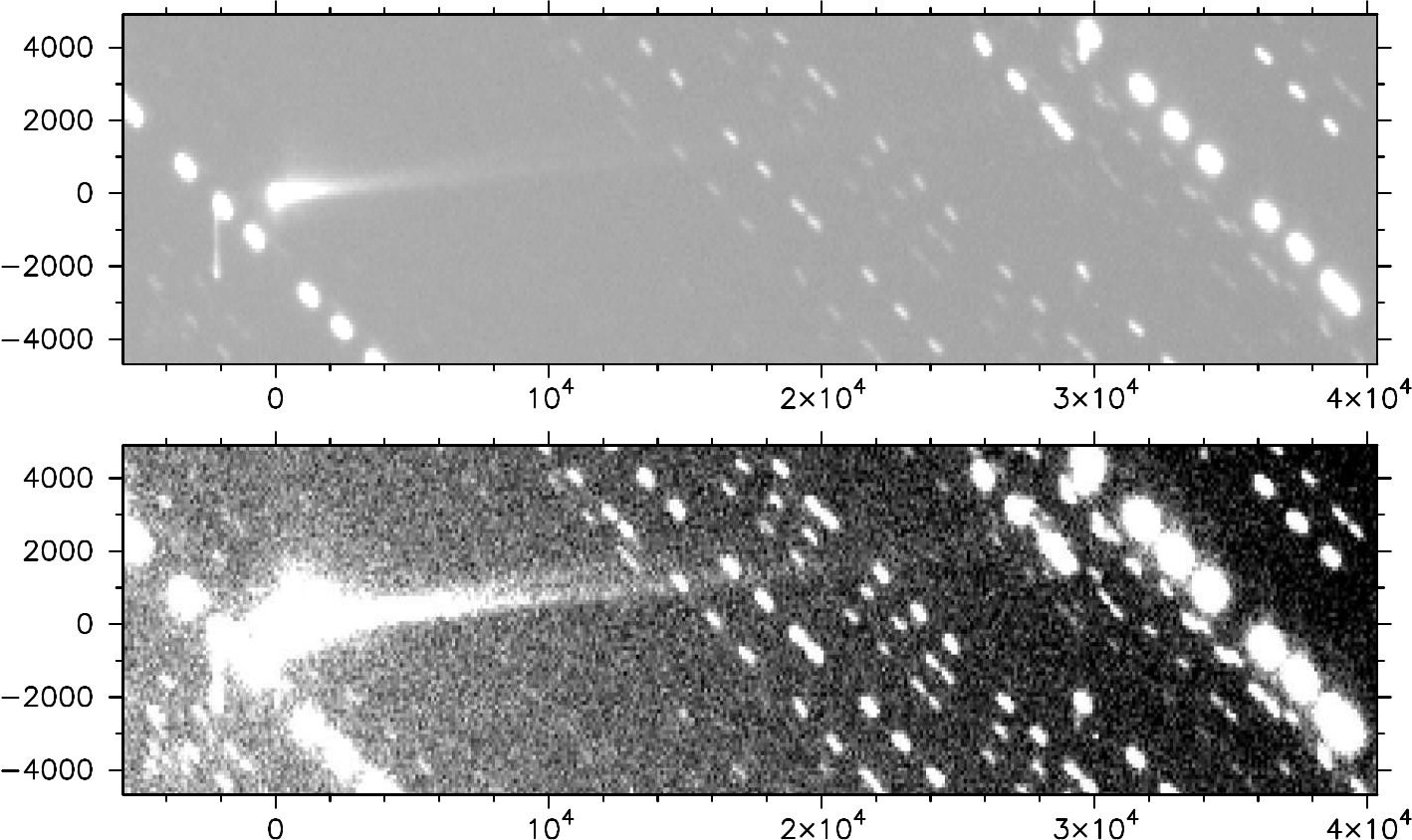}   
\caption{Appearance of the ejecta tail on an image taken at the 1-m telescope of the LULIN Observatory in Taiwan, on October 12, 18:19 UT (Z.-Y. Lin et al., 2023, submitted). The upper panel is the original, unstretched image, while the lower panel shows a heavily stretched display to show the double tail structure.  Axes are labeled in km projected on the sky plane. North is up, East to the left.
\label{fig:lulin}}
\end{figure}

The model image showing the double tail in comparison with the SPACEOBS observation is given in more detail in Figure \ref{fig:doubletail}. For purposes of comparison only, an additional image, taken at LULIN observatory 1m aperture telescope in Taiwan on October 12th, i.e., four days before the SPACEOBS image on October 16th, is also displaying the feature, with slightly higher signal-to-noise ratio than the SPACEOBS image in Figure \ref{fig:doubletail}, see Z.-Y. Lin et al. (2023, submitted) (Figure \ref{fig:lulin}). By November 2nd, image (d) in Table \ref{tab:SPACEOBS}, the secondary tail is not seen anymore, although it is still present  in the simulations, forming a very small angle with the main tail, when the synthetic image is shown heavily stretched (see Figure \ref{fig:spaceobs2}). This is not an effect of tail vanishing, but a geometric effect of the two synchrones associated to the main and secondary events getting more and more overlapped as the Earth becomes closer to the asteroid orbital plane (see PlAng column in Table \ref{tab:SPACEOBS}). 

A summary of the obtained dust parameters for this simple Monte Carlo model, describing the properties of each ejecta component (slow, fast, and late) is given in Table \ref{tab:models}.  

\begin{figure}[h]
\includegraphics[angle=-90,width=0.99\columnwidth]{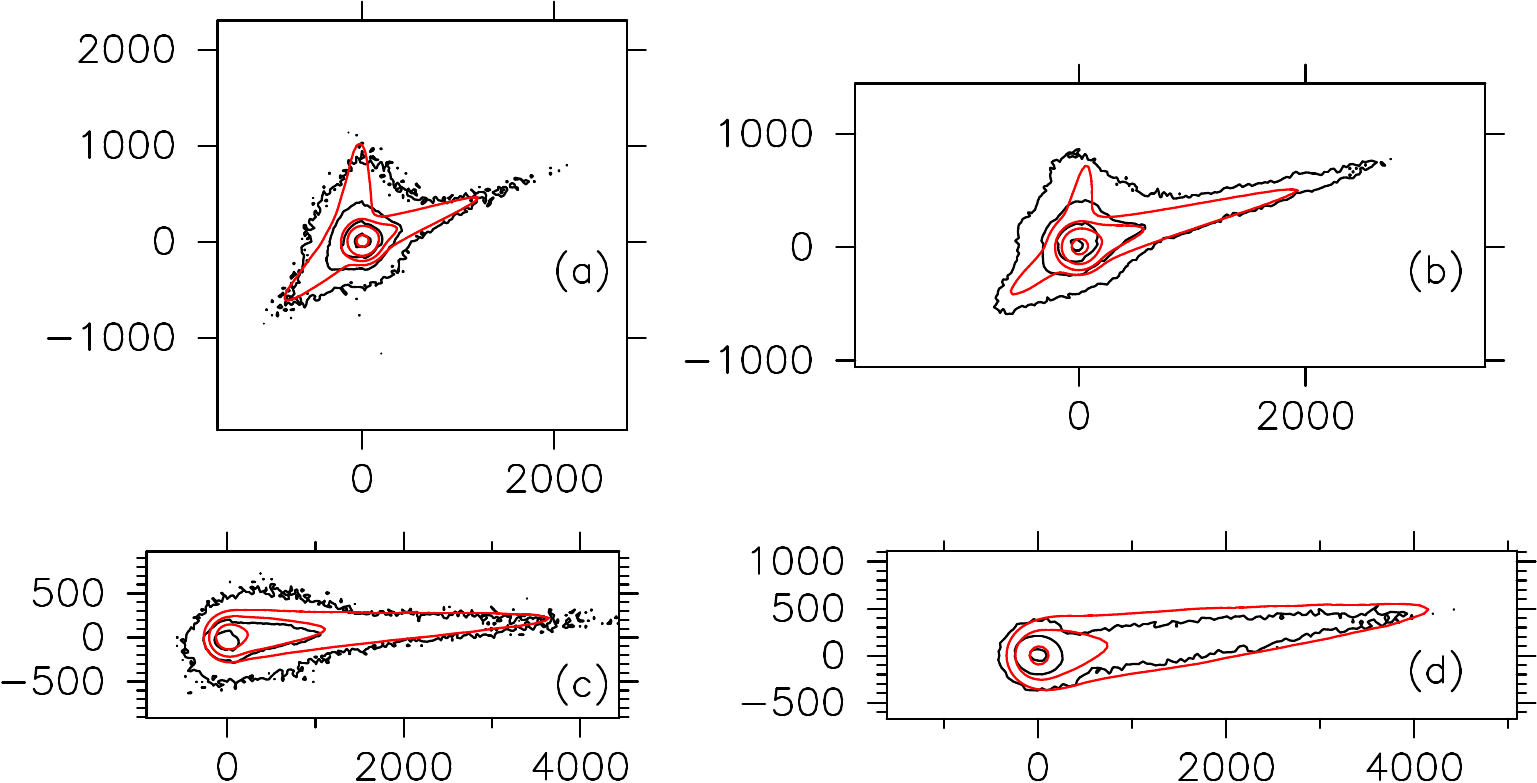}   
\caption{Isophote fields in the innermost regions for the earliest images (a), (b), (c) ,and (d) (see \ref{tab:SPACEOBS}). The observations are represented by black contours, and the model by red contours. The isophote levels are 17, 19, 21, and 23 mag arsec$^{-2}$ for images (a) and (b), and 17, 19, and 21 mag arsec$^{-2}$ for images (c) and (d). Axes are 
 labeled in km projected on the sky plane. North is up, East to the left in all images. 
\label{fig:isophote1}}
\end{figure}

For the earliest SPACEOBS images, it is useful to build an isophote field to make a detailed comparison with the model. Thus, Figure \ref{fig:isophote1} displays a comparison of the observed and modeled isophotes in the innermost regions close to the maximum condensation. As it is seen, the model fits are quite satisfactory, mostly taken into account the large range in brightness displayed. 

\begin{figure}[h]
\includegraphics[angle=-90,width=0.99\columnwidth]{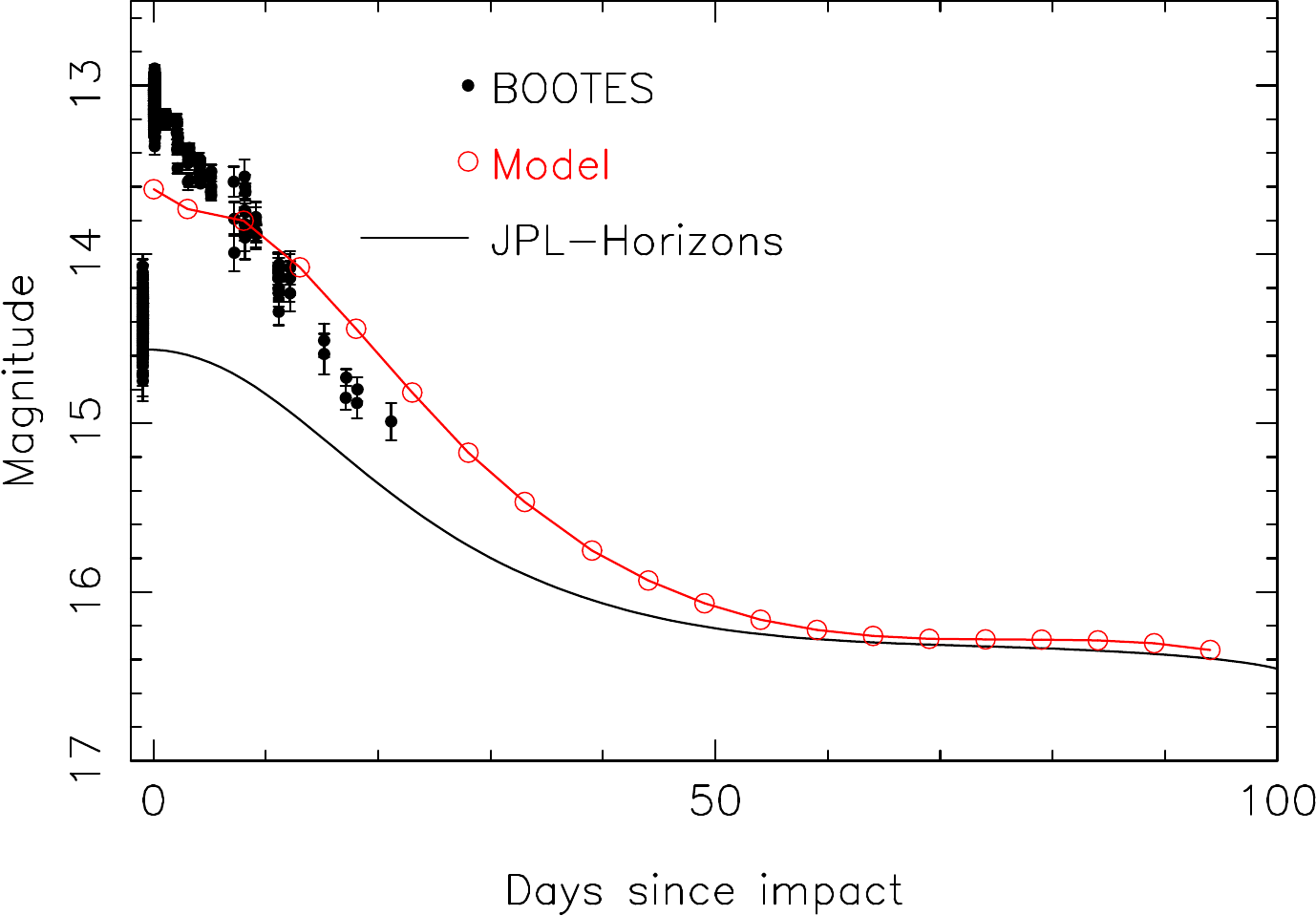}   
\caption{Aperture photometry by BOOTES compared with the simple Monte Carlo model predictions. The solid line is the magnitude as given in the JPL-Horizons system.  
\label{fig:BOOTES}}
\end{figure}

Finally, we compare the photometric measurements by BOOTES with the photometry calculated from this model. Figure \ref{fig:BOOTES} compares the magnitudes calculated with the model, at 5-day intervals since impact, and those by BOOTES. The model agrees reasonably well with the measurements. The JPL-Horizons apparent ``nuclear" magnitude data are also depicted for comparison at such epochs, which served also to check that the nuclear magnitudes computed with the model nucleus size, geometric albedo, and phase coefficient were correct. The JPL curve is obtained through the IAU H-G system magnitude model with absolute magnitude $H$=18.12 mag and $G$ parameter of $G$=0.15. Our model predicts a small relative maximum near the 8th day after impact, also hinted by the measurements.

\subsection{Detailed dynamical Monte Carlo modeling}\label{subsec:detailMC}

The visual inspection of the HST images reveals a complex dynamics in the neighborhood of the binary system. Owing to the superb spatial resolution ($\approx$2 km px$^{-1})$ that offers the HST WFC3 during the observational time span (see Table \ref{tab:HST}), a variety of structural details are seen in the images, as has been inventoried by \cite{2023Natur.616..452L}. The detailed dynamics of the particles close to the binary components can be described through the integration of the equation of motion of the individual particles subjected to the gravity fields of the two objects, as well as to the solar gravity and radiation pressure. We have already provided the input equations of the model \citep{2022MNRAS.515.2178M} when performing photometric predictions of the DART impact ejecta.  For completeness, we reproduce here the equation of motion of the ejected particles, where the reference frame has origin in the Didymos center of mass, and the two binary components are assumed as spherically-shaped: 
\begin{eqnarray}
\frac{\mathrm{d}^2\mathbf{r_d}}{\mathrm{d}t^2} = 
W_1\frac{\mathbf{r_d}}{r_d^3}
+W_2\frac{\mathbf{r_d}-\mathbf{r_s}}{\|\mathbf{r_d}-\mathbf{r_s}\|^3} \nonumber \\*
+ W_3\left[ \frac{\mathbf{r_s} - \mathbf{r_d}}{\|\mathbf{r_s} 
 - \mathbf{r_d}\|^3} - \frac{\mathbf{r_s}}{r_s^3}\right] + 
    W_4 \left[ \frac{\mathbf{r_{dsec}}}{r_{dsec}^3} -
    \frac{\mathbf{r_{sec}}}{r_{sec}^3} \right]  
\label{eqtraj}
\end{eqnarray}

\noindent
where $\mathbf{r_d}$ is the Didymos-to-dust
grain vector, $\mathbf{r_s}$ is the Didymos-to-Sun vector, 
$\mathbf{r_{dsec}}$ is the vector from the dust grain to Dimorphos,
and $\mathbf{r_{sec}}$ is the Didymos-to-Dimorphos
  vector. We have used the fact that
  $\mathbf{r_s}$=$\mathbf{r_d}$ + $\mathbf{r_{ds}}$, where
  $\mathbf{r_{ds}}$ is the vector from the dust grain to the
  Sun. Figure ~\ref{geometry} provides a schematic drawing of the
  vectors used.  The other terms are $W_1=-GM_P$, $W_3=GM_\odot$, and $W_4=GM_{sec}$, where $G$ is the gravitational constant, $M_P$ is the mass of Didymos, $M_{sec}$ is the mass of Dimorphos, and $M_\odot$ is the Sun mass. The remaining term, $W_2$,
  is given by: 

\begin{figure}[h]
\hspace{-1.0cm}
    \includegraphics[clip, trim=1.5cm 3.6cm 0.0cm 0.5cm,width=0.6\textwidth]{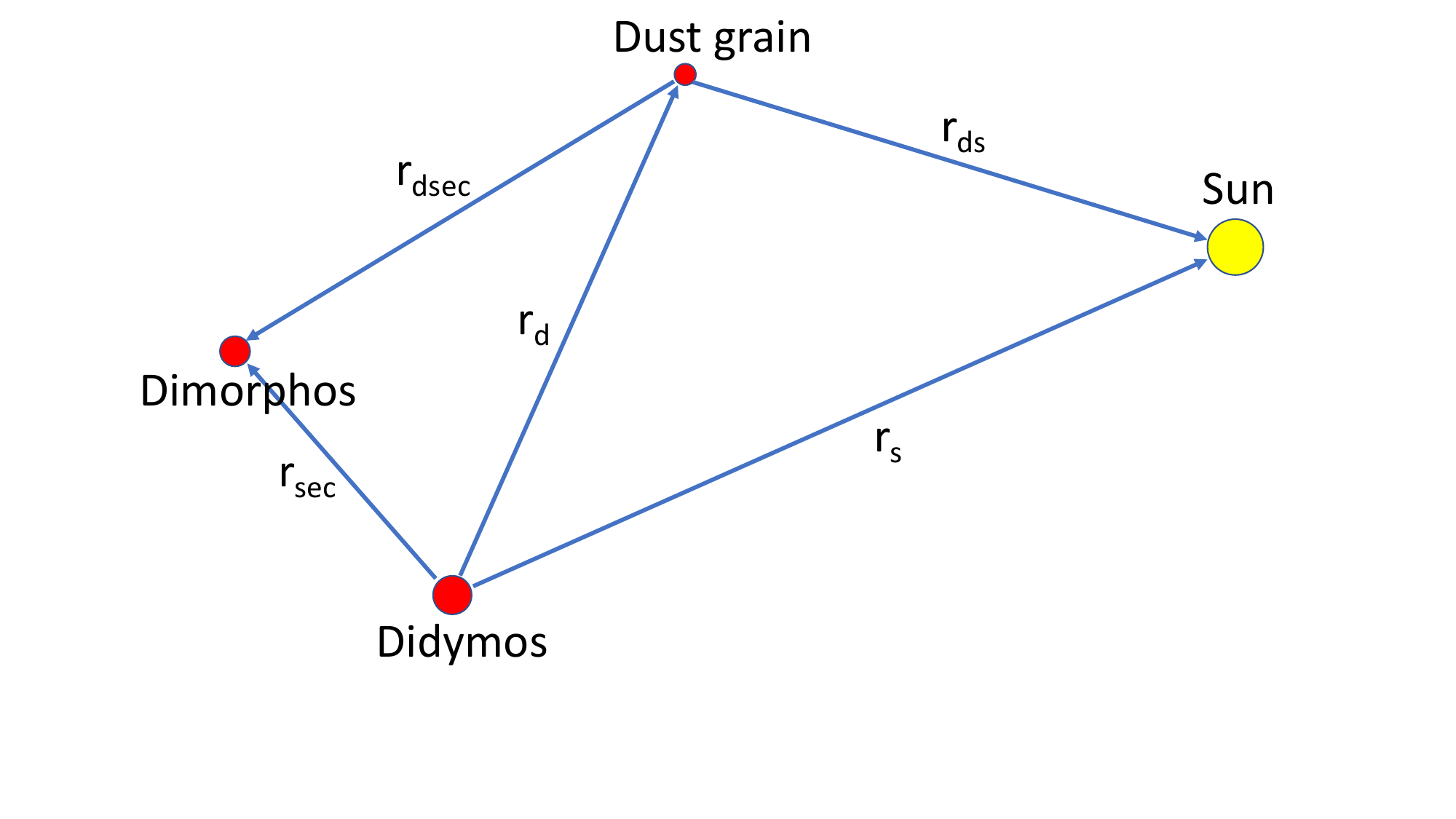}
    \caption{Schematic drawing of the vectors shown in Equation
       ~\ref{eqtraj}.}
    \label{geometry}
\end{figure}

\begin{equation}
  W_2 = \frac{Q_{pr}}{c} \frac{E_s}{4\pi} \frac{\pi d^2}{4 m_p}
\label{W2}
\end{equation}

\noindent
In Equation ~\ref{W2}, $c$ is the speed of light, $E_s$=3.93$\times$10$^{26}$ W is the total power radiated by the Sun, $d$ is the particle diameter ($d$=2$r$), and $m_p$ is the particle mass, $m_p=\rho_p (4/3)\pi r^3$.

As stated in \cite{2022MNRAS.515.2178M}, the model has been validated against the {\texttt{MERCURY}} N-body software package for orbital dynamics \citep{1999MNRAS.304..793C}. The initial conditions are assumed initially in a similar way as to the simple Monte Carlo (see section \ref{subsec:classMC}). However, in this more detailed model the larger particles might spend a significant time orbiting the neighborhood of the binary components until either collide with one of those bodies, or leave the system to interplanetary space \citep[see also][]{2022PSJ.....3..118R} Therefore the total mass ejected would actually be larger in this model than was found in the simple model since a fraction of that mass is lost in collision processes. For the same reason, the velocities of the particles will also have a somewhat different distribution. 

We modeled a subset of all the acquired HST images described in \cite{2023Natur.616..452L}. We used the same cone geometry as assumed for the ground-based image modeling, and the same size distribution parameters (the same broken power-law with the same limiting sizes). We begin by assuming similar values of the total masses ejected and the parameters associated to the velocity distribution. Then, we refine those parameters so as to give the best possible agreement between the model and the observations. As already stated, the fact that this model accounts for the orbital evolution in the neighborhood of the binary components implies necessarily a departure from the parameters obtained above from the relatively more simple model. As before, we considered a double speed ejecta component released immediately after the impact time, and a later secondary ejection event on October 2.5, 2022, the same date as in the simple Monte Carlo Model. In order to find a reasonable agreement with the evolution of the brightness and morphology of the observed features, the slow, hemispherical ejecta component, had a velocity simply given by $v=v_{esc}=0.09$ m s$^{-1}$  ($v_{esc}$ is the Dimorphos escape velocity) while the faster ejecta has $v$=0.225$\chi r^{-0.5}$ (where $\chi$ is a random number in the (0,1) interval). The two ejecta components contribute to the ejected mass in the same proportions as assumed in the Monte Carlo simple model. The ejecta speeds differ in factors of less than two relative to those found for the simple Monte Carlo model, which is quite reasonable taking into account the different approaches to the problem. 

Regarding the late emission, we assume the same parameters as in the simple Monte Carlo model approach, where this ejection event is characterized by isotropic emission, with particle speed $v=v_{esc}$. 

The total ejected mass (without taking into account the late emission) from this model is 6.4$\times$10$^6$ kg,  which would be close to the 4.2$\times$10$^6$ kg estimated using the simple Monte Carlo model considering that a significant fraction of the emitted mass in the detailed model is lost in collisions with either Didymos or Dimorphos. 
\begin{figure}[h]
\includegraphics[angle=-90,width=0.99\columnwidth]{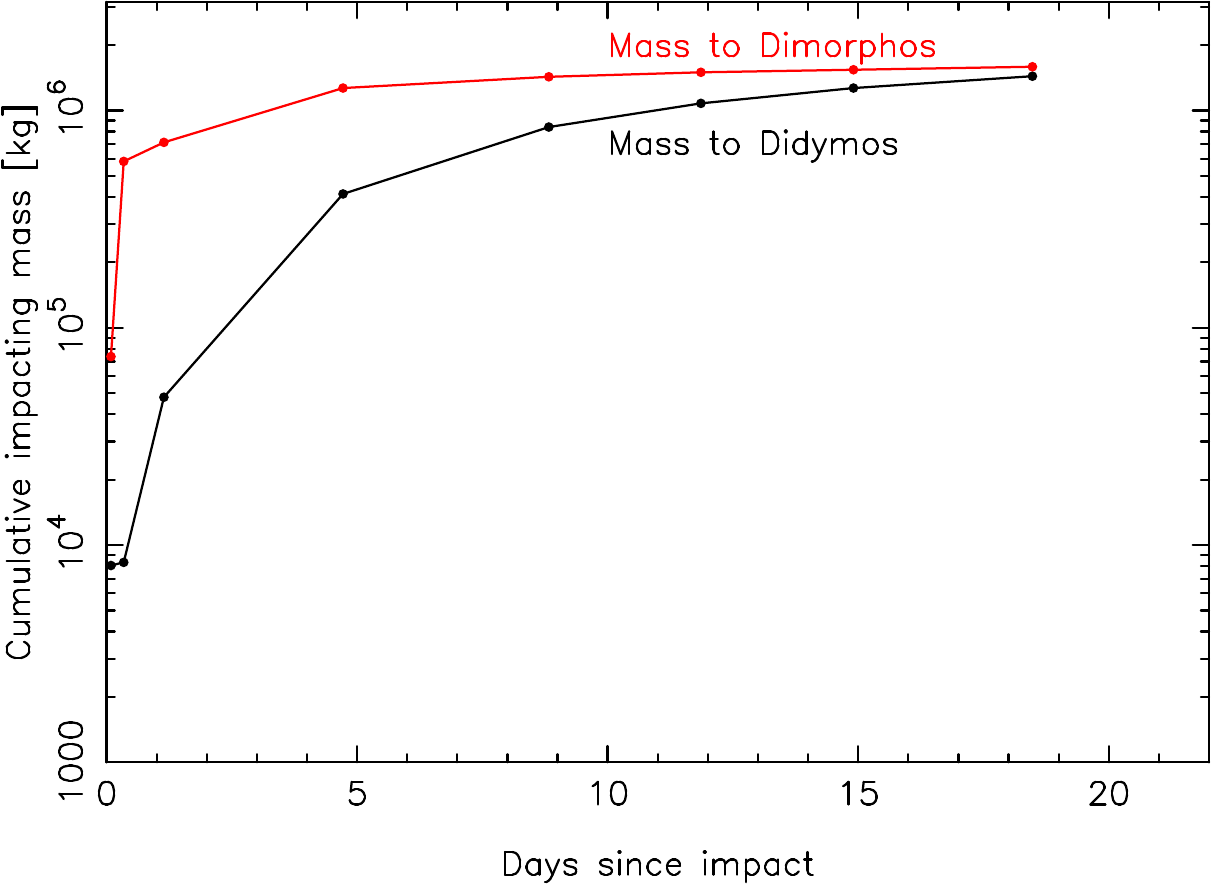}   
\caption{Cumulative impacting dust mass on Didymos and Dimorphos during the 20 days following impact. 
\label{fig:impactingmass}}
\end{figure}

\begin{figure}[h]
\includegraphics[angle=-90,width=0.99\columnwidth]{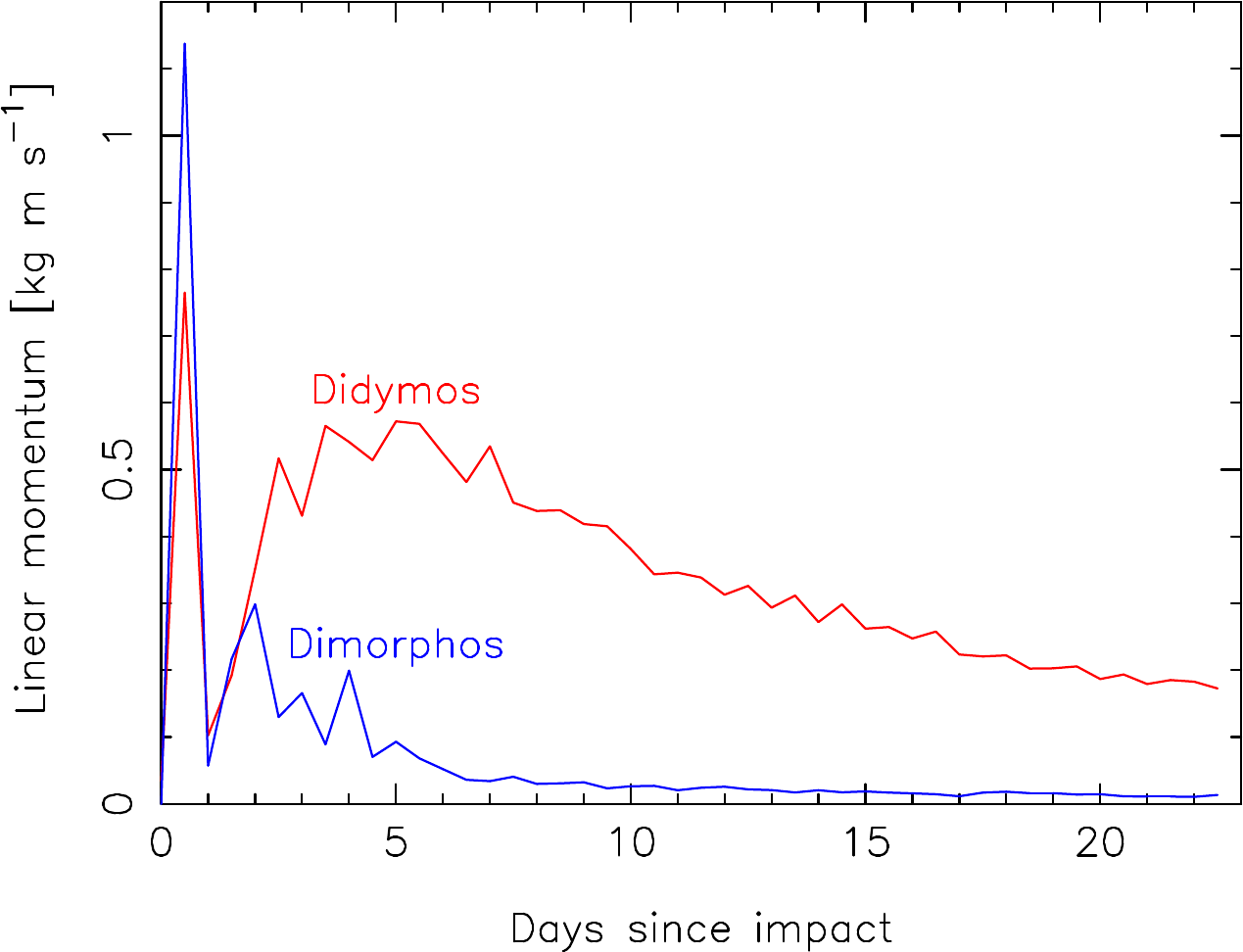}   
\caption{Linear momentum transferred, per second, to the surfaces of Didymos (red line) and Dimorphos (blue line), as a function of time since DART impact.   
\label{fig:impinging-momentum}}
\end{figure}

Concerning those colliding particles, Figure \ref{fig:impactingmass} depicts the cumulative dust mass impinging on those surfaces, and Figure \ref{fig:impinging-momentum} displays the total linear momentum transferred to the surfaces of Didymos and Dimorphos per unit time. From Figure \ref{fig:impactingmass} we see that for both surfaces, the total masses converge after some 20 days after impact to values close to 1.5$\times$10$^6$ kg. The momentum delivered to Dimorphos (Figure \ref{fig:impinging-momentum}) is higher than that on Didymos during the first few hours after impact, but the opposite occurs after $\approx$2 days, where the momentum on Didymos becomes dominant, reaching a maximum $\approx$5 days after impact. This behavior is confirmed by what is found by  \cite{2022PSJ.....3..118R}, where cm-sized particles were integrated. The momentum delivered to Didymos becomes dominant several days after the impact up to a time span of tens of days, thanks to particles that evolve within the binary system after being ejected from the impact crater. The re-impact velocities against Didymos is also higher with respect to the
ones against Dimorphos, reaching up to 80 cm/s in the cases analyzed in  \cite{2022PSJ.....3..118R}. We speculate that this momentum transfer could be the cause of, or at least contribute to, the generation of the secondary tail, as it peaks near the right time, but this argument would need further modeling about the effects of low-speed impacts on the surfaces of such bodies, which is beyond the scope of this work. In this regard, it should be noted that the secondary tail could also be associated to the dynamical evolution of particles that evolved for a while in the system and were then escaping after a few orbits, as revealed by the detailed dynamical analysis performed by Ferrari et al. (2023, in preparation).

The total ejected mass agrees very well with that estimated with the simple Monte Carlo model, as the total emission of 6.4$\times$10$^6$ kg of particles would actually be reduced to 3.4$\times$10$^6$ kg to unbound dust, because $\approx$3$\times$10$^6$ kg are lost in collisions with the binary components (see Figure \ref{fig:impactingmass}), leaving close to the 4.2$\times$10$^6$ kg of the simple Monte Carlo. 

\begin{figure}[ht!]
\includegraphics[angle=-90,width=0.99\columnwidth]{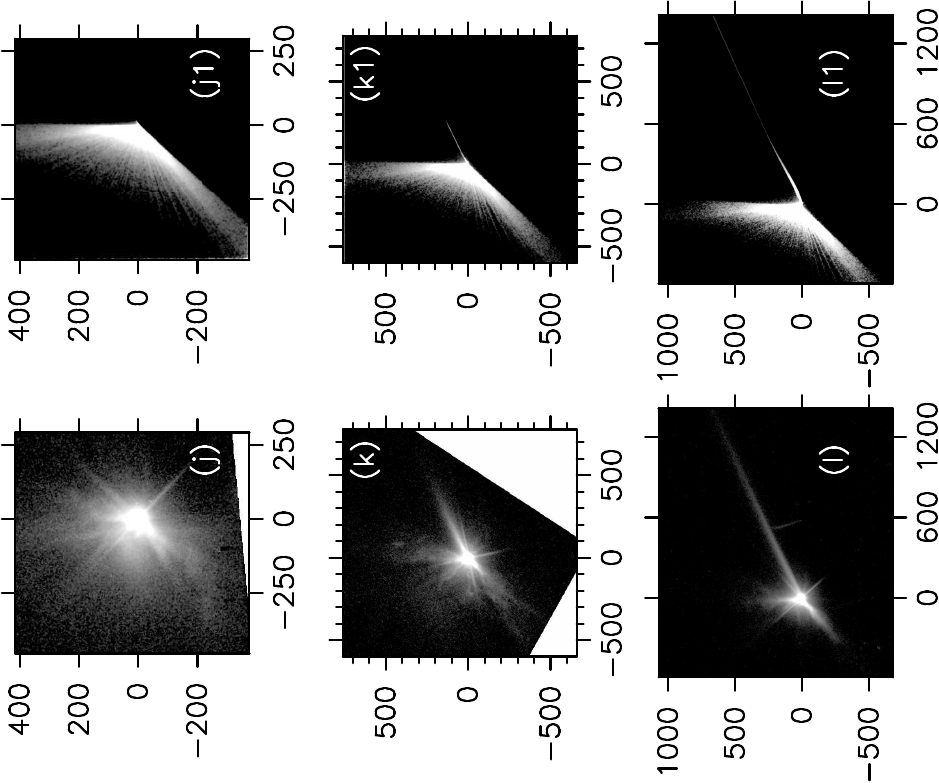}   
\caption{Panels (j), (k), and (l) display HST images at the corresponding dates in Table \ref{tab:HST}, and  panels (j1), (k1), and (l1) the corresponding synthetic images generated with the detailed Monte Carlo model described in \ref{subsec:detailMC}. The HST images are stretched between 22 and 17 mag arcsec$^{-2}$ and the modeled ones between 25 and 20 mag arcsec$^{-2}$ to show the antisolar tail that would not be seen otherwise because of their thinness. Axes are 
 labeled in km projected on the sky plane. North is up, East to the left in all images.
\label{fig:HST1}}
\end{figure}

\begin{figure}[ht!]
\includegraphics[angle=-90,width=0.99\columnwidth]{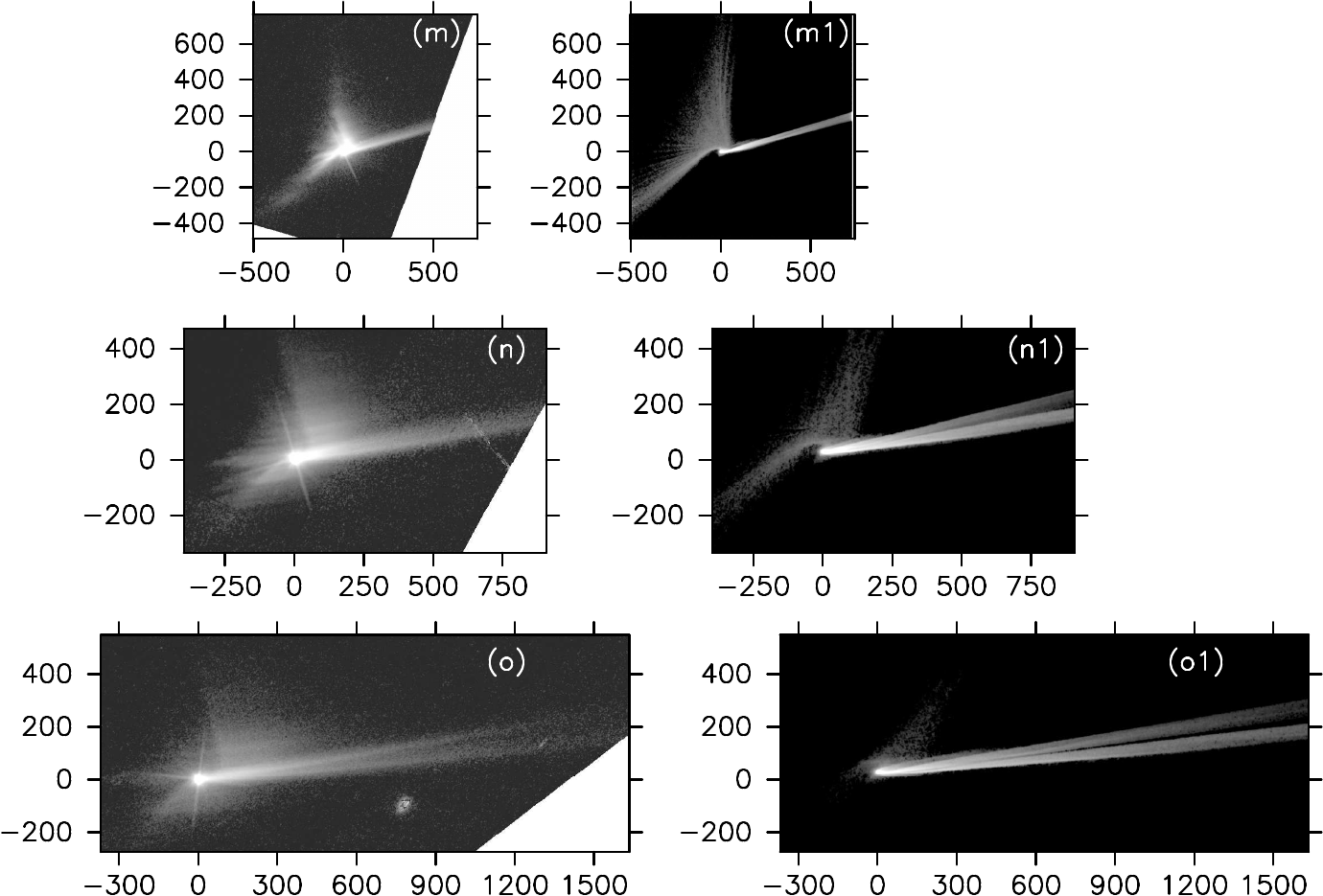}   
\caption{Panels (m), (n), and (o) display HST images at the corresponding dates in Table \ref{tab:HST}, and  panels (m1), (n1), and (o1) the corresponding synthetic images generated with the detailed Monte Carlo model described in \ref{subsec:detailMC}. All images are stretched between 22 and 17 mag arcsec$^{-2}$. Axes are labeled in km projected on the sky plane. North is up, East to the left in all images.
\label{fig:HST2}}
\end{figure}

\begin{figure}[ht!]
\includegraphics[angle=-90,width=0.99\columnwidth]{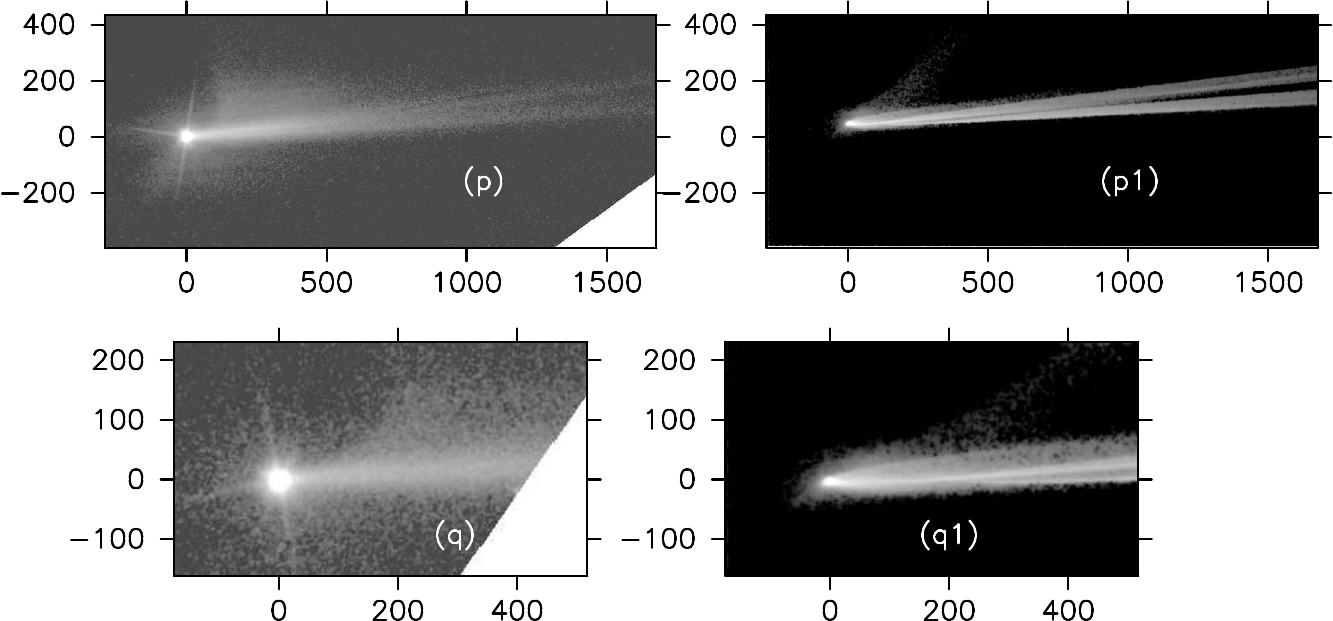}   
\caption{Panels (p), and (q), display HST images at the corresponding dates in Table \ref{tab:HST}, and  panels (p1), (q1), and (l1) the corresponding synthetic images generated with the detailed Monte Carlo model described in \ref{subsec:detailMC}. All images are stretched between 22 and 17 mag arcsec$^{-2}$. Axes are labeled in km projected on the sky plane. North is up, East to the left in all images.
\label{fig:HST3}}
\end{figure}

\begin{figure}[ht!]
\includegraphics[angle=-90,width=0.99\columnwidth]{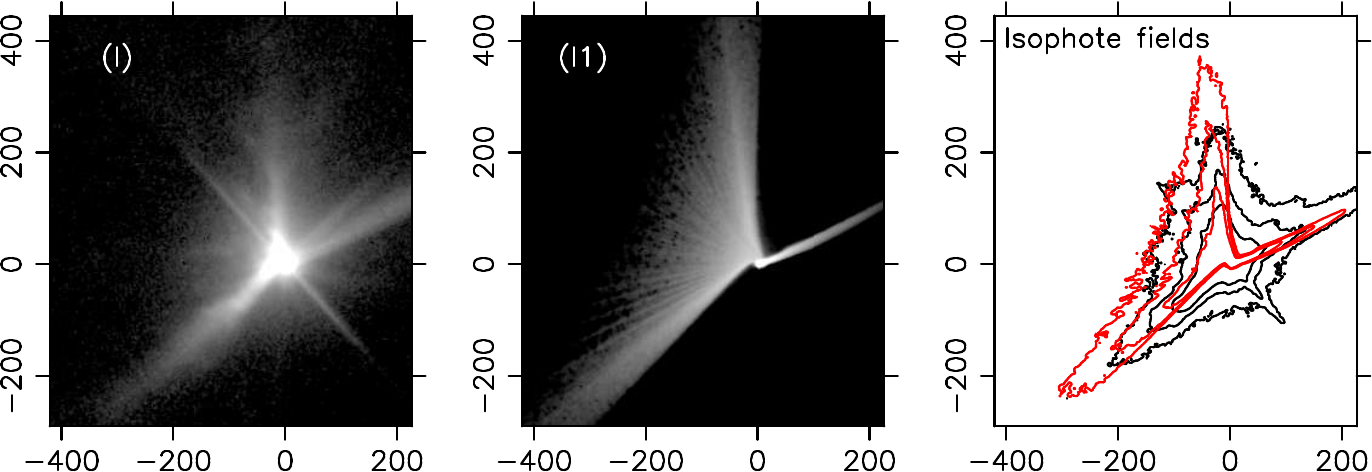}   
\caption{Panel (l) displays the central portion of image (l) (see Table \ref{tab:HST}) and panel (l1), displays the corresponding modeled image. The rightmost panel depicts the isophote fields with contours on 20, 19, and 18 mag arsec$^{-2}$ (black contours correspond to the observation, and red contours to the model). All panels are labeled in km projected on the sky, and are oriented North up, East to the left. 
\label{fig:HST_isophotes}}
\end{figure}

For purposes of comparison, the synthetic images generated are convolved with the HST point-spread function for the appropriate filter and camera used. The HST observations along with the model images are then shown in Figures \ref{fig:HST1}, \ref{fig:HST2}, and \ref{fig:HST3}. In Figure \ref{fig:HST1}, the modeled images are heavily stretched to display the antisolar tail extent, which otherwise would be unseen owing to their thinness. The central portion of image (l) and its model (l1), along with the isophote fields are depicted in Figure \ref{fig:HST_isophotes}. The outermost modeled isophotes associated to the conical emission are in line with the observations, although the antisolar tail becomes too narrow in comparison with what it is observed. On the other hand,  the length of the tail constrains the minimum particle radius to $\approx$1$\upmu$m, as assumed in the simple Monte Carlo model: a larger minimum size would result in a too short tail, and a smaller minimum size would display an incipient tail already on image (j). The secondary tail, which competes in brightness with the main tail, clearly appears in the images (o) and (p) (see Figures \ref{fig:HST2}, and \ref{fig:HST3}). This feature had to be modeled by an ejecting dust mass of the order of half of the main event, i.e., $\approx$3$\times$10$^6$ kg. However an important fraction of that mass is re-impacting again Didymos and/or Dimorphos, so that only about half of that late mass is released to the interplanetary medium, i.e., 1.5$\times$10$^6$ kg.  This mass is about twice of that estimated with the simple Monte Carlo, but since these HST images are far better in resolution than those obtained with SPACEOBS, we prefer to rely on this value. Considering all the ejected masses, the total contribution to the mass in unbound orbits becomes 4.9$\times$10$^6$ kg, which agrees fairly well with the value estimated from the simple Monte Carlo model (4.2$\times$10$^6$ kg). A summary of the dust properties from the detailed model is given in Table \ref{tab:models}.  

\begin{figure}[ht!]
\includegraphics[angle=-90,width=0.99\columnwidth]{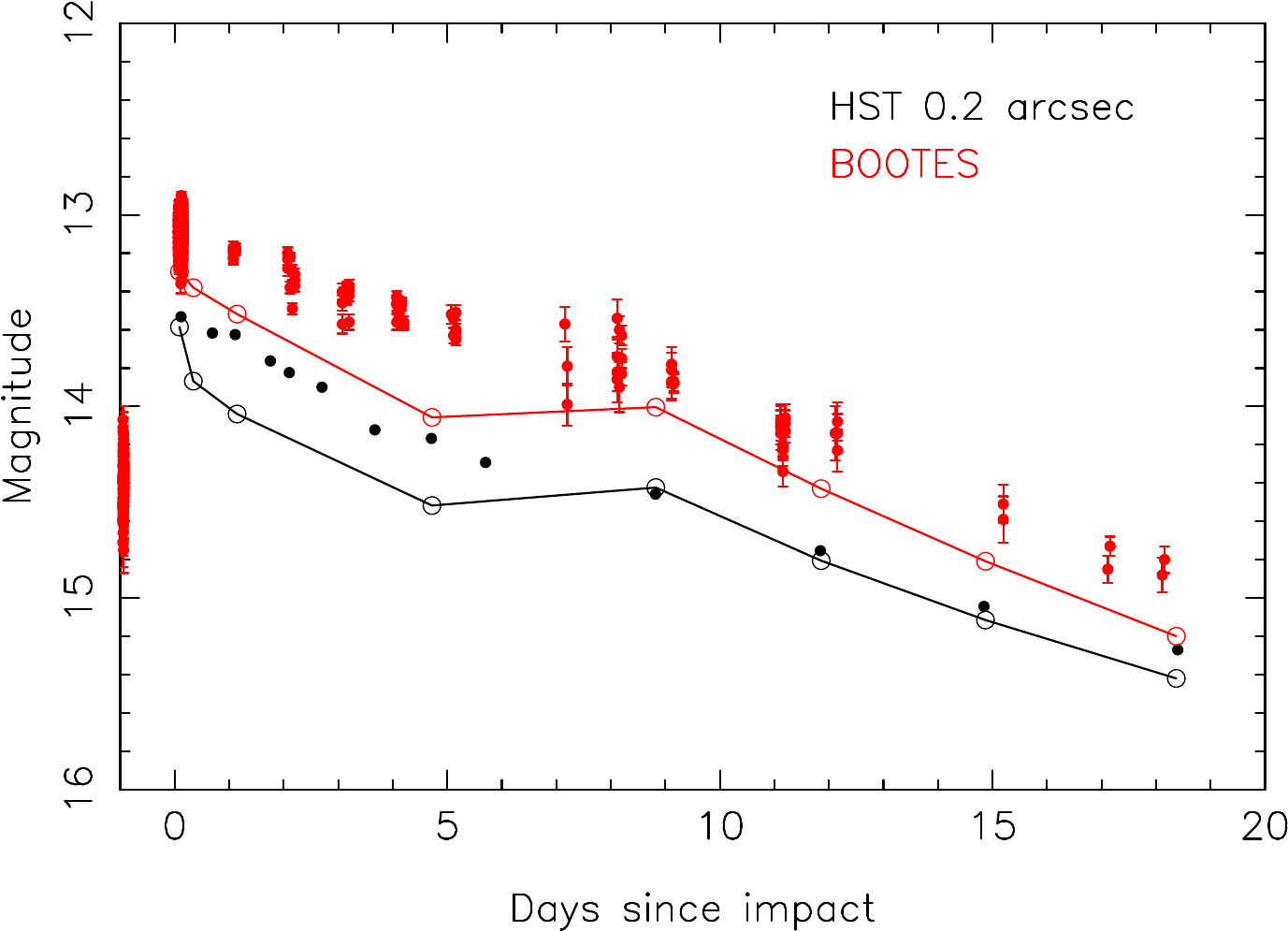}   
\caption{Lightcurves from HST images with 0.2\arcsec aperture \citep{2023Natur.616..452L} (black solid circles) compared with BOOTES photometry (with 6\arcsec-7\arcsec aperture depending on seeing conditions) (red solid circles) and the model results (open circles connected by black and red solid lines for HST and BOOTES, respectively).
\label{fig:HST-BOOTES}}
\end{figure}

The photometric measurements on the HST images using a 0.2\arcsec aperture \citep{2023Natur.616..452L} are displayed in Figure \ref{fig:HST-BOOTES}, together with BOOTES photometry. These two lightcurves are consistent, showing a constant difference of 0.5$\pm$0.1 magnitude, owing to the different apertures used. The model results for the subset of HST images shown in Table \ref{tab:HST} are found to be in line with the measurements for both data sets.  

Although many of the observed features are captured in the model, some of them remain unexplained. Thus, in the early images, the observed tail is broader than the modeled ones, as we have already illustrated in Figure \ref{fig:HST_isophotes}. The tail thickness is mostly influenced by the ejection speeds, so that setting a larger ejection speed we would get a thicker tail. However, model runs show that while this would work for the earlier images, this would lead to much broader and more diffuse tails than observed for the later images of Figures \ref{fig:HST2} and \ref{fig:HST3}. On the other hand, while the southern ejecta curtain shows a temporal evolution similar to that retrieved with the model, the northern branch of
the ejecta (diffuse ejecta in Figure \ref{fig:features}) shows a different evolution, showing a North-East orientation, while the model predicts a  North-West orientation instead, following the direction of the radiation pressure force.  In addition, there are also
some features in the sunward direction that the model
does not reproduce, likely due to an asymmetric ejecta cone  \citep{2023acm..Hirabayashi}. 
The origin of such discrepancies is unclear, and many physical parameters are likely contributing. Note that the model uses spherical particles moving in the gravitational fields of assumed spherical bodies, which is not the case. The dynamics of non-spherical particles in asteroidal or cometary environments  is certainly more complex \citep{2017Icar..282..333I, 2017..Ferrari,
2022MNRAS.510.5142M,2023acm..Ivanosvki}, as well as the gravitational fields of non-spherical bodies. In addition, the model assumes that the ejected particles have constant mass and size, excluding any disruption and/or fragmentation processes during their motion. In addition, the ejection pattern is much more complex than described by a simple conical geometry, showing an intrincate structure with non-radial filaments 
\citep{2023Natur.616..457C,2023acm..Dotto}. Future models should incorporate those effects to see how they affect the resulting dust structures in an attempt to understand the physical processes involved. 

\section{Comparison with active asteroids}

Active asteroids constitute a recently discovered class of objects in the Solar System which, having typical asteroidal orbits, exhibit comet-like appearance \citep[e.g.][]{2022arXiv220301397J}. 
The DART mission has resulted in the artificial activation of one of such objects, as anticipated by \cite{2023acm..Tancredi}, and, following their suggestion, it is interesting to briefly compare the DART results with those observed in naturally activated asteroids. A more extended comparison of the DART results with the natural active asteroids will be the subject of a forthcoming paper (Tancredi et al., in preparation). Following the discovery of 133P/Elst-Pizarro in 1996 
\citep{2004AJ....127.2997H}, some 40 active asteroids have been detected so far \citep{2022arXiv220301397J}. Amongst the physical mechanisms that can trigger activation, are ice sublimation, rotational destabilization, and the result of an impact, or a combination of those. The sample of impacted objects is statistically not significant, as out of that active asteroids population, only four objects have been identified to be most likely activated by an impact with another body, namely 354P/LINEAR \citep[e.g.][]{2012A&A...537A..69H,2013ApJ...769...46A}, (493) Griseldis \citep{2015DPS....4741403T,2022arXiv220301397J}, P/2016 G1 (PANSTARRS) 
\citep{2016ApJ...826L..22M, 2017AJ....154..248M, 2019ApJ...877L..41M,2019A&A...628A..48H}, and the large asteroid (596) Scheila \citep{2011ApJ...740L..11I,2011ApJ...738..130M}.  In the case of the DART impact, most of the physical parameters are known in advance: the masses of the impactor and the impacted bodies, the relative velocity, the geometry of the impact, and the impact time. For the natural impacts, none of those parameters are known, not even the collision speed: the relative speed between impactor and impacted body in the DART collision is $\approx$6 km s$^{-1}$, which is similar to the mean collision speed in the main asteroid belt ($\sim$5 km s$^{-1}$), but as pointed out by 
\cite{2011SSRv..163...41O}, asteroids indeed experience a significant fraction of impacts at velocities much smaller or larger than the "canonical" value. In addition, those naturally impacted objects are normally found already active during dedicated sky surveys programs such as Pan-STARRS , LINEAR, or Catalina Sky Survey, and the impact occurred at an unspecified earlier time. In consequence, the evolution of the dust structures cannot be systematically followed as in the DART collision event, where dedicated ground-based and space telescope campaigns were planned well in advance. Therefore, for cases other than DART, retrieving information on the dust parameters from scarce data, normally a few images and/or spectra during a short time window, becomes difficult. The determination of the impact time is made through Finson-Probstein or Monte Carlo modeling of the observed tails, but it is always difficult to assess owing to the complex morphology of the observed dust patterns, and the large amount of dust physical parameters involved.  In a way similar to the findings in DART impact, the resulting particle ejection speeds of the observed ejecta are found to be very small, of the order of the escape velocities of the impacted bodies \citep[e.g.][and references therein]{2021MNRAS.506.1733M}, confirming the scaling laws predictions that most of the mass is ejected at low speeds \citep[e.g.][]{1983JGR....88.2485H}.  

In contrast, one of the most remarkable differences between those impacted asteroids and Dimorphos is the duration of the observed tails. The fading of the tail on those natural active asteroids commonly occurs in a time span of several weeks, while the DART tail is still observable after more than 9 months since impact \citep{2023acm..Li} Possibly the binary nature of the impacted object is playing a role in that long survivability of the tail in keeping relatively large particles orbiting the neighborhood of the binary components for a long time before being ejected to the interplanetary medium. A thoughtful analysis of the long-lasting tail is beyond the scope of this paper.

\section{Conclusions}\label{sec:conclusions}
The observed dust ejecta after the collision of the DART spacecraft with Dimorphos, the satellite of the (65803) Didymos system, has been modeled by Monte Carlo dust tail codes. The observations, taken from the Earth and from the HST, which has the advantage of exploring the ejecta behaviour at two different spatial resolutions and spatial scales, are analyzed by simple and detailed Monte Carlo modeling. From the ground-based data, and using our simple Monte Carlo model, we conclude that the differential size distribution of the particles could be represented by a broken power-law function with index $\kappa$=--2.5 for particles between 1 $\upmu$m and 3 mm, and with index $\kappa$=--3.7 for particles of radii between 3 mm and 5 cm. The ejecta pattern might be explained, on one hand by particles being ejected along the wall of a hollow cone with axis pointing to RA=130$^\circ$, DEC=17$^\circ$, with relatively large speeds, and on the other hand, by particles emitted hemispherically at Dimorphos escape speed, oriented in the same way as the emitting cone. With this configuration, and ejected mass of approximately 6$\times$10$^6$ kg, most of the observed features can be reproduced both morphologically and photometrically, at all the epochs included in the present analysis, keeping in mind that this estimate is always a lower limit, as the presence of large boulders in the distribution, having large mass, but contributing negligibly to the brightness, cannot be excluded. The detailed Monte Carlo model takes into account the rigorous motion of the particles in the neighborhood of the binary system. With this model, various details observed in the ejecta on the HST images have been reproduced, although there remains some that so far cannot be captured with such a model. In any case, the model parameters used to explain the ground-based images can also explain the detailed structures seen in the HST images. In particular the northern and southeastern streams associated to the hollow cone emission, the early evolution of the ejecta, the length of the anti-sunward tail, and the double tail pattern. The northern component of the double tail could be associated to reimpacting material on Didymos, as the momentum carried by the impacting particles peaks at nearly the same epoch as that needed to generate the secondary tail. However, further modeling is clearly needed to test this conclusion. There are also many structures seen in the HST images, such us the northern diffuse pattern, unreproducible with the model, that needs further modeling including additional processes, such as particle collisions, fragmentation, and disruption phenomena.

\section{Acknowledgments}

We are very grateful to the two anonymous referees for their careful reviewing of the manuscript, and their detailed suggestions, which have helped us to improve considerably the manuscript. 

Some of the data presented in this paper were obtained from the Mikulski Archive for Space Telescopes (MAST) at the Space Telescope Science Institute. The specific observations analyzed can be accessed via \dataset[doi:10.17909/pvc8-fk24]{http://dx.doi.org/10.17909/pvc8-fk24}

FM acknowledges financial supports from grants PID2021-123370OB-I00, 
P18-RT-1854 (Junta de Andalucia), and from the Severo Ochoa grant CEX2021-001131-S funded by MCIN/AEI/10.13039/501100011033.

ACB acknowledges funding by the NEO-MAPP project, grant agreement 870377,  EC H2020-SPACE-2018-2020 / H2020-SPACE-2019.

AJCT acknowledges support from Spanish MICINN project PID2020-118491GB-I00 and the support of the technical staff at INTA-CEDEA where the BOOTES-1 station is located.

JLO ackcnowledges support from contract PID2020-112789GB-I00.


%

\vspace{5mm}
\facilities{HST(STScI), SPACEOBS, BOOTES Global Network, LULIN}

\bibliography{sample631}{}
\bibliographystyle{aasjournal}



\end{document}